\documentclass[a4paper,useAMS,usenatbib]{mnras}
\usepackage{graphics,epsfig,psfig}
\usepackage[normalem]{ulem}
\usepackage{xcolor}
\usepackage{float}
\usepackage{subfig}
\usepackage{graphicx}
\usepackage[]{inputenc,amssymb}
\usepackage{natbib}
\usepackage{url}
\usepackage{widetext}
\usepackage{amsmath}
\usepackage{cancel}

\usepackage[style=base, singlelinecheck=off, font=small, labelfont=bf, justification=raggedright]{caption}
\def \be{\begin{equation}}
\def \ee{\end{equation}}
\def \bea{\begin{eqnarray}}
\def \eea{\end{eqnarray}}

\def\apj{ApJ}
\def\mnras{MNRAS}
\def\prd{PRD}
\def\aap{A\&A}
\definecolor{webgreen}{rgb}{0,.5,0}
\definecolor{webbrown}{rgb}{.6,0,0}

\def\aap{A\&A}
\def\apj{ApJ}
\def\apjs{ApJS}
\def\apjl{ApJL}
\def\mnras{MNRAS}

\def\nat{Nature}

\def\physrep{Physics Reports}         
\def\prd{Phys.Rev.D}         

\def\jcap{JCAP}

\title[Weak lensing of gravitational waves]{Probing the theory of gravity with gravitational lensing of gravitational waves and galaxy surveys}
\author[Mukherjee, Wandelt, \& Silk]{Suvodip Mukherjee$^{1,2}$\thanks{mukherje@iap.fr}, Benjamin D. Wandelt$^{1,2,3}$ \thanks{wandelt@iap.fr} \& Joseph Silk$^{1,2,4,5}$ \thanks{silk@iap.fr}\\
$^{1}$ Institut d'Astrophysique de Paris,  98bis Boulevard Arago, 75014 Paris, France\\
$^{2}$ Sorbonne Universites, Institut Lagrange de Paris,  98 bis Boulevard Arago, 75014 Paris, France\\
$^{3}$ Center for Computational Astrophysics, Flatiron Institute, 162 5th Avenue, 10010, New York, NY, USA\\
$^{4}$ The Johns Hopkins University, Department of Physics \& Astronomy, 3400 N. Charles Street, Baltimore, MD 21218, USA\\
$^{5}$ Beecroft Institute for Cosmology and Particle Astrophysics, University of Oxford, Keble Road, Oxford OX1 3RH, UK\\
}
\pubyear{2019}
\begin{document}
\label{firstpage}
\pagerange{\pageref{firstpage}--\pageref{lastpage}}

\maketitle
\begin{abstract}
The cross-correlation of gravitational wave strain with upcoming galaxy surveys probe theories of gravity in a new way. This method enables  testing the theory of gravity by combining the effects from both gravitational lensing of gravitational waves and the propagation of gravitational waves in spacetime. We find that within 10 years, the combination of the  Advanced-LIGO and VIRGO detector networks with planned galaxy surveys should detect weak gravitational lensing of gravitational waves in the low redshift Universe ($z<0.5$). With the next generation gravitational wave experiments such as Voyager, LISA, Cosmic-Explorer and Einstein Telescope, we can extend this test of the theory of gravity to larger redshifts by exploiting the synergies between  electromagnetic wave and gravitational wave probes. 
\end{abstract}
\section{Introduction}
The Lambda-Cold Dark Matter (LCDM) model of cosmology matches well all of the best available data sets probing cosmic microwave background (CMB), galaxy distributions, and supernovae \citep{Aghanim:2018eyx, Alam:2016hwk, Abbott:2017wau,Abbott:2018wog,Jones:2018vts}. However  $95\%$ of the content of this  model  is composed of dark energy and dark matter that is not yet understood. Several theoretical models aid our  understanding of  this unknown part of the Universe and lead to observable predictions  which can be explored using electromagnetic probes of the Universe. With the discovery of gravitational waves \citet{PhysRevLett.116.061102, PhysRevLett.119.161101}, a  new observational probe has opened up which is capable of bringing new insights to our studies of  the Universe. Study of the imprints of the dark matter and late-time cosmic acceleration on gravitational waves will provide  new avenues for  understanding these unknowns. 

With the ongoing ground-based gravitational wave observatories such as Advanced-LIGO (Laser Interferometer Gravitational-Wave Observatory) \citep{Martynov:2016fzi}, Virgo (Virgo interferometer)\citep{TheVirgo:2014hva}, and with the upcoming observatories such as KAGRA (Large-scale Cryogenic Gravitational wave Telescope) \citep{Akutsu:2018axf} and LIGO-India \citep{Unnikrishnan:2013qwa}, we can measure the gravitational wave signals from binary neutron stars (NS-NS), stellar-mass binary black holes (BH-BH), black hole-neutron star (BH-NS) systems over the frequency range $10-1000$ Hz and up to a redshift of about one. Along with the ground-based gravitational wave detectors, space-based gravitational wave experiment such as LISA (Laser Interferometer Space Antenna) \citet{PhysRevD.93.024003, 2017arXiv170200786A} are going to measure the gravitational wave signal emitted from supermassive binary black holes in the frequency range $\sim 10^{-4}- 10^{-1}$ Hz. In the future, with ground-based gravitational wave experiments such as Voyager, Cosmic Explorer \citet{Evans:2016mbw} and the Einstein Telescope\footnote{\url{http://www.et-gw.eu}}, we will be able to probe gravitational wave sources up to a redshift $z=8$ and $z=100$ respectively \citet{Sathyaprakash:2019nnu}. This exquisite multi-frequency probe of the Universe is going to provide an avenue towards understanding the history of the Universe up to high redshift and unveil the true nature of dark energy and dark matter.  

In this paper, we explore the imprints of the cosmic density field on astrophysical gravitational waves and forecasts for measuring them using cross-correlations with the galaxy density field and galaxy weak lensing. Intervening cosmic structures induce distortions, namely magnifications and demagnifications in the gravitational wave signal when gravitational waves propagate through  spacetime. According to the theory of general relativity, these distortions should be universal for gravitational wave signals of any frequency for a fixed density field. These imprints should exhibit correlations with other probes of the cosmic density field such as galaxy surveys, as both of the distortions have a common source of origin. 

We estimate the feasibility of measuring the lensing of the gravitational wave signals emitted from astrophysical sources in the frequency band $10-1000$ Hz and $10^{-4}- 10^{-1}$ Hz by cross-correlating with the upcoming large-scale structure surveys such as DESI (Dark Energy Spectroscopic Instrument) \citet{Aghamousa:2016zmz}, EUCLID \citet{2010arXiv1001.0061R}, LSST (The Large Synoptic Survey Telescope) \citet{2009arXiv0912.0201L}, SPHEREx \citet{Dore:2018kgp}, WFIRST (Wide Field Infrared Survey Telescope)\citep{Dore:2018smn}, etc. which also probes the cosmic density field. The cross-correlation of the gravitational wave signal with the cosmic density field is going to validate the theory of gravity, nature of dark energy and the impact of dark matter on gravitational wave with a high SNR over a large range of redshift as discussed in Sec. \ref{forecast}.  {The auto-correlation of the GW sources can also be used as a probe of the lensing signal. However this is going to be important in the regime of large numbers of GW sources with high statistical significance such as from the Einstein Telescope \footnote{\url{http://www.et-gw.eu}} and Cosmic Explorer \citep{Evans:2016mbw}.}

The paper is organized as follows. In Sec. \ref{weaklensing}, we discuss the effect of lensing on the galaxy distribution and on gravitational waves. In Sec. \ref{implication_gw}, we discuss the implication of the gravitational lensing of gravitational waves in validating the theory of gravity and the standard model of cosmology. In Sec. \ref{estimator} and Sec. \ref{forecast}, we obtain the estimator for measuring the signal and the forecast of measuring this from the upcoming gravitational wave observatories and LSS missions. Finally, in Sec. \ref{conclusion}, we discuss the conclusions of this work and its future scope.
 
\section{Weak lensing}\label{weaklensing}
The general theory of relativity predicts  universal null geodesics for both gravitational wave and electromagnetic wave, that can be written in terms of the Friedmann-Lemaitre-Robertson-Walker (FLRW) metric with line elements given by $ds^2= -dt^2 + a(t)^2(dx^2+ dy^2+ dz^2)$, where $a(t)$ is the scale factor, is set to one at present and decreases to zero in the past. However, in the presence of  density fluctuations, both electromagnetic and  gravitational waves propagate along  the perturbed FLRW metric which can be written as $ds^2= -(1+2\Phi) dt^2+ a(t)^2(1+2\Psi) (dx^2+ dy^2+ dz^2)$. 
One of the inevitable effects of matter perturbations on both   electromagnetic  and gravitational waves is gravitational lensing which can lead to magnification, spatial deflections and time delays in the arrival times of the signal \citep{ 1993ApJ...404..441K, Kaiser:1996tp, 2001PhR...340..291B}. In this paper, we discuss the aspect of  gravitational weak lensing due to the distribution of the density fluctuations in the Universe and its effect on  cosmological observables such as the galaxy field and gravitational waves. 

\subsection{Weak lensing of galaxies}
The observed distribution of the galaxies can be translated into a density field as a function of  sky position and redshift of the source as
 \begin{align}\label{gal}
\begin{split}
\delta (\hat n, z)= \frac{{\rho}_m(\hat n, z)}{\bar {\rho}_m (z)} -1,
\end{split}
\end{align}
where $\rho_{m} (\hat n, z)$ is the matter density at a position $\hat n$ at the redshift $z$ and $\bar \rho_{m}$ denotes the mean density at $z$. The density field of the galaxies is assumed to be a biased tracer of the underlying dark matter density field which can be written in Fourier space in terms of the scale-dependent bias $b_g(k)$ as  $\delta_g (k)= b_g (k) \delta_{DM} (k)$. In the Fourier domain, we  define the galaxy power spectrum in terms of the dark matter power spectrum by the relation $P_g(k)= b_g (k)^2 P_{DM}(k)$, where the power spectrum is defined as $\langle \delta_{X'}(k) \delta^*_{X}(k')\rangle= (2\pi)^3 P_{X'X} (k) \delta_{D} (\vec k- \vec k')$ ($\delta_D$ denotes the Dirac delta function).

The inhomogeneous distribution of matter between galaxies and the observer causes distortions in  galaxy shapes. The lensing convergence field can be written in terms of the projected line of sight matter density  as  
\begin{align}\label{kappa-gal}
\begin{split}
\kappa_{g} (\hat n) = &\int_{0}^{z_s} d z \frac{3}{2}\frac{\Omega_m H_0^2(1+z)\chi(z)}{cH(z)}\\&\int_z^\infty dz' \frac{dn_{gal}(z')}{dz'}\frac{(\chi(z')-\chi(z))}{(\chi(z'))} \delta(\chi(z)\hat n, z),
\end{split}
\end{align}
where $z_s$ is the redshift of the source plane and $\frac{dn_{gal}(z')}{dz'}$ is the normalized redshift distribution of galaxies. The convergence field is related to the cosmic shear by the relation \citet{Kaiser:1996tp}
\begin{align}\label{kappa-shear}
\begin{split}
\kappa_{g} (\vec k) = \frac{k^2_1 - k^2_2}{k^2_1 + k^2_2} \gamma_1(\vec k)+ \frac{k_1k_2}{k^2_1 + k^2_2} \gamma_2(\vec k),
\end{split}
\end{align}
where, $\gamma_1$ and $\gamma_2$ are the cosmic shears in the Fourier space with $\vec k= (k_1, k_2)$. From the observables such as the galaxy distributions and its shape, the lensing field can be inferred from the data of the upcoming galaxy surveys \citep{Aghamousa:2016zmz,2010arXiv1001.0061R,2009arXiv0912.0201L,Dore:2018smn,Dore:2018kgp}.

\subsection{Weak lensing of gravitational wave}
The strain of the gravitational waves from  coalescing binaries can be written in terms of the redshifted chirp mass $\mathcal{M}_z= (1+z)\mathcal{M}$ as \citep{1987thyg.book.....H, Cutler:1994ys,Poisson:1995ef,maggiore2008gravitational}
\begin{align}\label{gw-1}
\begin{split}
h (f_z)&= \mathcal{Q}(\text{angles})\sqrt{\frac{5}{24}}\frac{G^{5/6}\mathcal{M}_z^2 (f_z\mathcal{M}_z)^{-7/6}}{c^{3/2}\pi^{2/3}d_L} e^{i\phi_z},
\end{split}
\end{align}
  where $\mathcal{M}= \frac{(m_1m_2)^{3/5}}{(m_1+m_2)^{1/5}}$ is the true chirp mass in terms of the masses of the individual binaries $m_1$ and $m_2$ and $d_L$ is the luminosity distance to the binaries which can be written for the standard LCDM model of cosmology as  $d_L= (c(1+z)/H_0)\int_0^z \frac{dz'}{\sqrt{\Omega_m(1+z')^3 + (1-\Omega_m)}}$.  {$G$ and $c$ are the gravitational constant and speed of light respectively}. Propagation of the gravitational wave in the presence of a matter distribution leads to magnification in the 
strain of gravitational waves  {in the geometric optics limit} which can be written as \footnote{ {The effects of matter perturbations on the GW signal apart from the effect from lensing are explored in previous studies \citep{Laguna:2009re}}}  \citep{Takahashi:2005ug, Laguna:2009re, Cutler:2009qv,  Camera:2013xfa, Bertacca:2017vod} 
\begin{align}\label{gw-lensing}
\begin{split}
\tilde h(\hat n, f_z)=h(f_z)[1 + \kappa_{gw} (\hat n)],
\end{split}
\end{align}
where $\kappa_{gw} (\hat n)$ is the lensing  {convergence} and can be written in terms of the  density fluctuations $\delta(\chi(z)\hat n, z)$ along the line of sight by the relation
\begin{align}\label{kappa-1}
\begin{split}
\kappa_{gw} (\hat n) = &\int_{0}^{z_s} d z \frac{3}{2}\frac{\Omega_m H_0^2(1+z)\chi(z)}{cH(z)}\\&\int_z^\infty dz' \frac{dn_{gw}(z')}{dz'}\frac{(\chi(z')-\chi(z))}{(\chi(z'))} \delta(\chi(z)\hat n, z),
\end{split}
\end{align}
where $\frac{dn_{gw}(z')}{dz'}$ is the normalized redshift distribution of the gravitational wave sources.  {The redshift distribution of the GW sources can be determined from the electromagnetic 
counterparts or else can also be estimated by using the clustering of the GW source positions with the galaxy catalogs \citep{PhysRevD.93.083511, Mukherjee:2018ebj}}. According to the theory of general relativity and the standard model of cosmology, in the absence of anisotropic stress, metric perturbations are related by  $\Phi=-\Psi$, and are related to the matter perturbation by the relation   
\begin{align}\label{pois-1}
\begin{split}
\nabla^2\Phi= 4\pi G a^2\rho_m \delta,
\end{split}
\end{align}
where $G$ is the gravitational constant and $\rho_m$ is the matter density. 

\section{Probe of general theory of relativity and dark energy}\label{implication_gw}
Gravitational waves bring a new window to validate the general theory of relativity and cosmological constant as the correct explanation of the theory of gravity and cosmic acceleration. Propagation of gravitational waves in the presence of cosmic structures brings two avenues to explore the Universe:
\begin{enumerate}
    \item  the propagation of gravitational wave in the spacetime is defined by a perturbed FLRW metric 
 \begin{align}\label{GW-prop}
\begin{split}
\mathcal{D}_\mu\mathcal{D}^\mu h_{\alpha\beta} + 2R_{\mu\alpha\nu\beta}h^{\mu\nu} =0. 
\end{split}
\end{align} 
where $\mathcal{D}_\mu$ is the covariant derivative and $R_{\mu\alpha\nu\beta}$ is the curvature tensor; and
\item  For a perfect fluid energy-momentum tensor, the evolution of the metric perturbations $\Phi$ and $\Psi$ and its relationship with the matter perturbation can be written at all cosmological epochs by the relation \citet{Hu:2007pj}
 \begin{align}\label{struc_1}
\begin{split}
g (k,z)\equiv  \frac{\Psi (k,z) +\Phi(k,z)}{\Psi(k,z)-\Phi(k,z)}=& 0, \\
\frac{1}{2}\nabla^2 (\Psi (k,z)-\Phi (k,z))=& 4\pi Ga^2\rho_m \delta,
\end{split}
\end{align}
\end{enumerate}

Gravitational lensing of gravitational waves is going to probe both of these aspects of the theory of gravity from the observations.   
Several alternative theories of gravity consider deviations in Eq. \eqref{GW-prop} \citep{Cardoso:2002pa, Saltas:2014dha, Nishizawa:2017nef} and Eq. \eqref{struc_1} \citep{Carroll:2006jn, Bean:2006up,Hu:2007pj,Schmidt:2008hc, Silvestri:2013ne, Baker:2014zba, Lombriser:2015sxa, Lombriser:2016yzn, Saltas:2014dha, Baker:2015bva, Slosar:2019flp} such as (i) $g (k,z)\neq 0$, (ii) the modified Poisson Equation, (iii) running of the Planck mass which acts like a frictional term leading to damping of the gravitational strain more than the usual damping due to the luminosity distance, (iv) mass of the graviton (v) speed of propagation of gravitational waves and (vi) the anisotropic stress term as a source term in the gravitational wave propagation equation Eq. \eqref{GW-prop}. A joint analysis of all of these observational aspects can be performed to test the validity of the general theory of relativity and cosmic acceleration using the framework discussed in this paper. 

An avenue to study gravitational wave propagation in the presence of matter perturbations is to  cross-correlate the gravitational wave strain with the cosmic density field probed by other avenues such as the galaxy field, CMB and line-intensity mapping. In this paper, we explore the cross-correlation of the galaxy field with the gravitational wave strain from astrophysical gravitational wave sources which are going to be measured from  upcoming gravitational wave observatories such as Advanced-LIGO \citet{Evans:2016mbw}, Virgo \citet{TheVirgo:2014hva}, KAGRA \citet{Akutsu:2018axf}, LIGO-India \citet{Unnikrishnan:2013qwa} and LISA \citet{PhysRevD.93.024003, 2017arXiv170200786A}. The joint study of the gravitational wave data along with the data from 
large-scale structure and CMB missions will probe all observable signatures (listed above) due to modifications in the theory of gravity and dark energy by measuring the distortion in the gravitational wave strain.

 {Previous studies have discussed the effect of gravitational lensing of gravitational waves on measuring  cosmological parameters using  gravitational wave events only \cite{Takahashi:2005ug, Laguna:2009re, Cutler:2009qv,  Camera:2013xfa, Bertacca:2017vod,Congedo:2018wfn}.  {The measurement of the luminosity distance using  gravitational waves can also probe the curvature of the Universe \citep{Congedo:2018wfn}.} In this work, we examine the cross-correlation of the gravitational wave strain with  electromagnetic probes. This approach introduces two new aspects. First, we can study the synergy between the electromagnetic sector and the gravitational wave sector. Different theories of gravity lead to modifications in Eq. \eqref{GW-prop} and Eq. \eqref{struc_1}, and a joint estimation of both of these aspects is necessary to constrain these models. Second, the cross-correlation study with large samples of data from galaxy surveys makes it possible to detect the lensing signal from noise-dominated gravitational wave data.}

\section{Measuring the lensing of the gravitational wave using cross-correlations with galaxy surveys}\label{estimator}
\subsection{Theoretical signal}
The propagation of the electromagnetic waves and gravitational waves along the same perturbed geodesics brings an inevitable correlation between the galaxy ellipticity and distortion in the gravitational wave strain according to the theory of general relativity and the standard LCDM model of cosmology.  {The gravitational wave strain gets lensed by the foreground galaxies present between the gravitational wave source and us. The theoretical power spectrum of the gravitational wave lensing in the spherical harmonic basis can be written as \footnote{ {We have used the Limber approximation ($k= (l+1/2)/\chi(z)$)}.}}
 \begin{align}\label{gal-gw}
\begin{split}
C^{\kappa_{gw}\kappa^j_{gal}}_l \equiv& \langle (\kappa_{gw})_{lm} (\kappa_{g})_{l'm'} \rangle \delta^K_{ll'}\delta^K_{mm'}\\=& \int \frac{d z}{\chi^2} \frac{H(z)}{c} \bigg[W_{\kappa_{gw}}(\chi(z))W^j_{\kappa_{g}}(\chi(z)) \\& \times P_{\delta}((l+1/2)/\chi(z))(\chi(z)) \bigg],\\
C^{\kappa_{gw}\delta^j}_l \equiv&\langle (\kappa_{gw})_{lm} \delta_{l'm'} \rangle \delta^K_{ll'}\delta^K_{mm'} \\ =&\int \frac{d z}{\chi^2} \frac{H(z)}{c} b_g \bigg[W_{\kappa_{gw}}(\chi(z))W^j_{\delta}(\chi(z)) \\& \times P_{\delta}((l+1/2)/\chi(z)) (\chi(z))\bigg],
\end{split}
\end{align}
where  $W^i_{\kappa_{gw}},\,W^i_{\kappa_{g}}$ and $W^i_{\delta}$ are the kernels for gravitational wave lensing, galaxy lensing and the galaxy field respectively for the tomographic bins indicated by the bin index $j$ in redshift, which can be defined as 
 \begin{align}\label{gal-gw}
\begin{split}
W_{\kappa_{gw}/\kappa_{gal}} (\chi(z))=& \frac{1}{H(z)}\int_z^\infty dz' \frac{dn_{gw/ gal}(z')}{dz'}\frac{(\chi(z')-\chi(z))}{(\chi(z'))},\\
W_{\delta}^j (\chi(z))=& \frac{dn}{dz}(z_j).
\end{split}
\end{align}

\begin{figure*}
\centering
\subfloat[]{\includegraphics[trim={0cm 0.cm 0cm 0.cm},clip,width=1\textwidth]{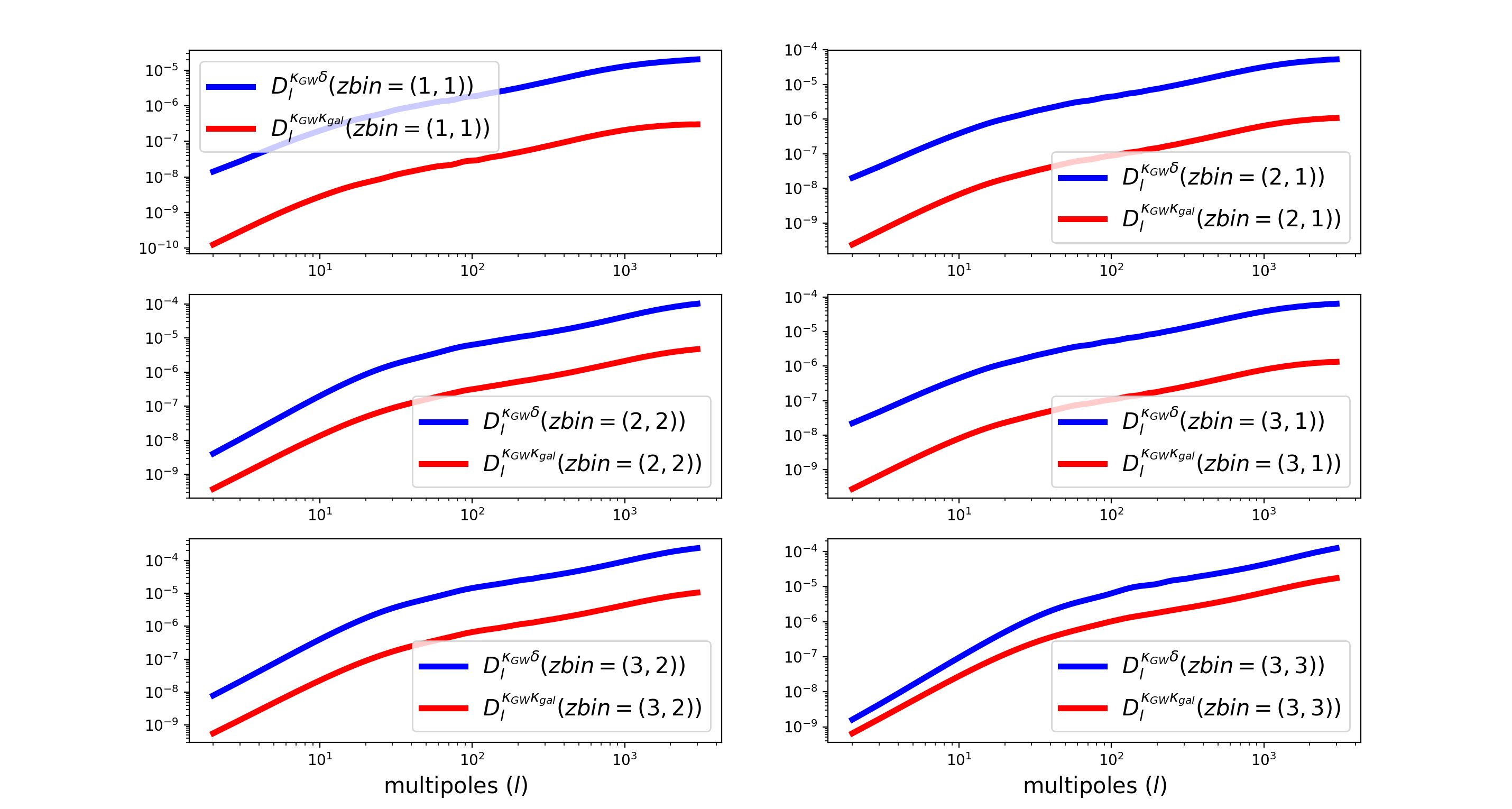}}\\
\subfloat[]{\includegraphics[trim={0cm 0.cm 0cm 0.cm},clip,width=1.\textwidth]{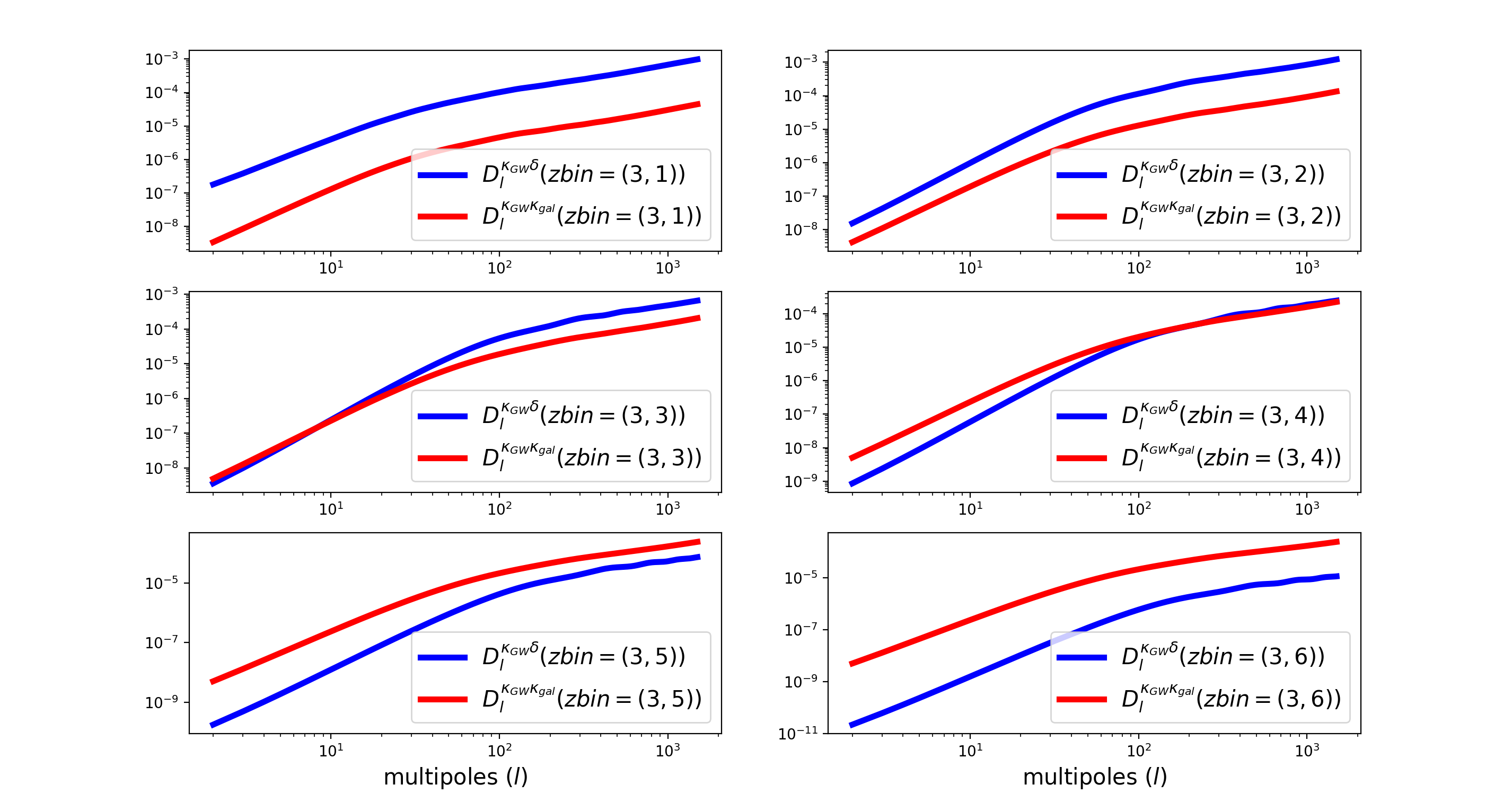}}\\
\caption{We show the correlation for (a) the LIGO sources and (b) the LISA sources between $\kappa_{gw}$ with $\kappa_{g}$ and $\delta$ for different tomographic redshift bins indicated by the $z_{bin}$ index given by $(z_{gw}, z_{g/\kappa_g})$. The redshift range is taken from $[0,1]$ and $[0,3]$ for the LIGO and LISA sources with the bin width $\Delta z=0.2$ and $\Delta z=0.5$ respectively. For the galaxy survey, we have taken the redshift distribution for (a) Model-I and (b) Model-II with $z_0=0.5$ and for gravitational wave we assume a single source plane $dn_{gw}/dz (z_s)= \delta_D(z-z_s)$. }
\label{fig:cls}
\end{figure*}

In this analysis, we have taken a linear bias model with the value of the bias parameter as $b_g=1.6$ \citep{2012MNRAS.427.3435A,2017MNRAS.470.2617A,Desjacques:2016bnm} and the non-linear dark matter power spectrum from the numerical code CLASS \citep{2011JCAP...07..034B, 2011arXiv1104.2932L}.  The power spectra for galaxy lensing and gravitational wave lensing, $C^{\kappa_{gw}\kappa^j_{gal}}_l$ and  $C^{\kappa_{gw}\delta^j}_l$, are plotted in  Fig. \ref{fig:cls} as a function of the redshift of the gravitational wave sources. The redshift distribution of the gravitational wave sources is taken as $\frac{dn_{gw}}{dz} (z_s)=\delta(z-z_s)$ and the distribution of the galaxies is taken according to Models  
  \begin{align}\label{dndzeq}
\begin{split}
\frac{dN}{dz} (z_s)=& z_s^2\exp(-(z_s/z_0)^{3/2}) \,\,\, \text{Model-I},\\
\frac{dN}{dz} (z_s)=& \bigg(\frac{(z_s/z_0)^2}{(2z_0)}\bigg)\exp(-z_s/z_0) \,\,\, \text{Model-II}.
\end{split}
\end{align}
The signal of the cross-correlation gets stronger at high redshift due to more overlap between the gravitational wave lensing kernel and the galaxy kernel. 

\begin{figure*}
    \centering
    \includegraphics[trim={0 0 0 0cm}, clip, width=1.\textwidth]{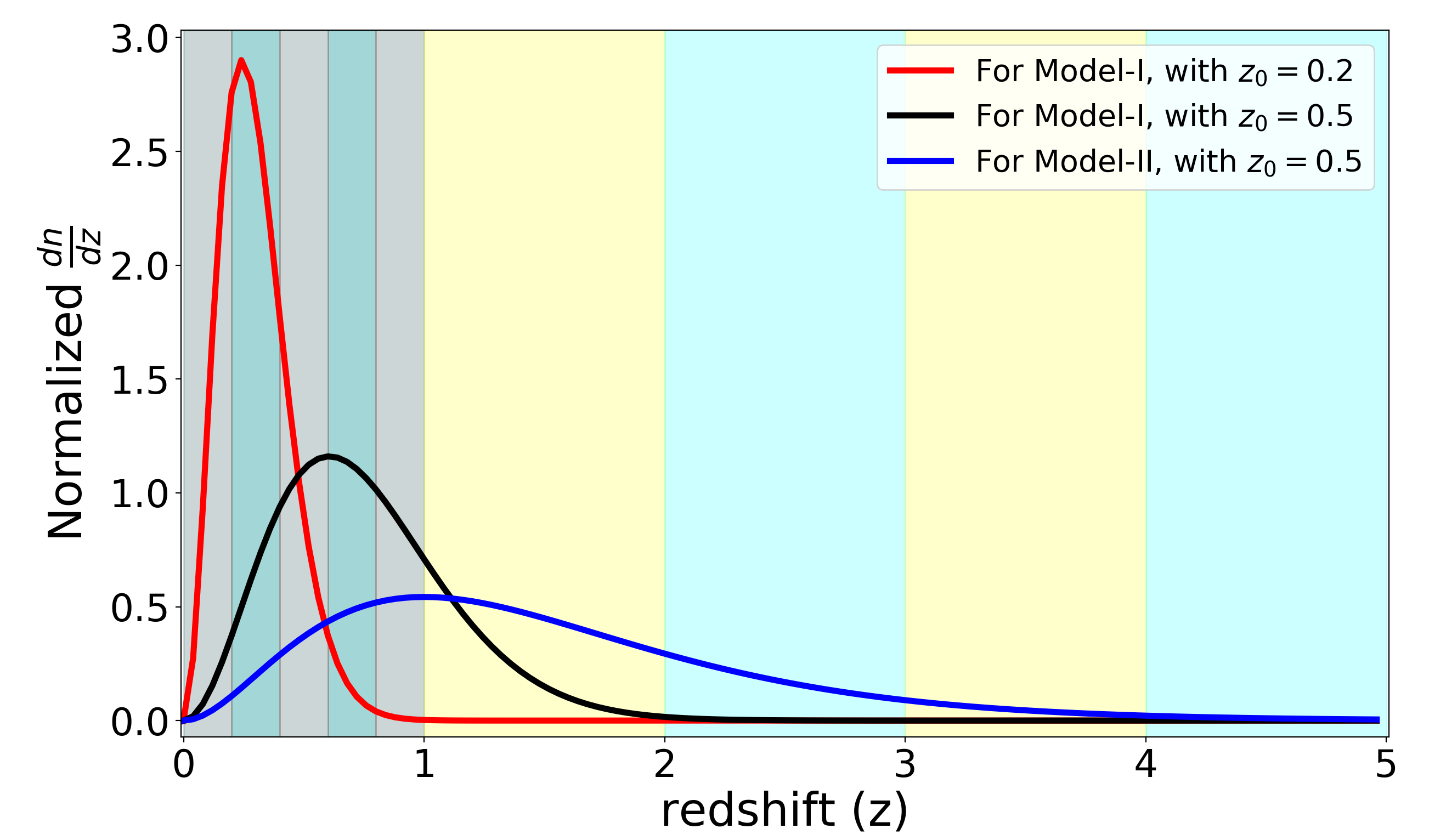}
    \caption{We show the normalized distribution of the galaxy surveys for Model-I and Model-II given in Eq. \ref{dndzeq}. We have used the case with Model-I and $z_0=0.2$ for the cross-correlation with the NS-NS and BH-NS sources which can be observed by LIGO. For the BH-BH sources from LIGO, we have used Model-I with $z_0=0.5$. For the LISA gravitational wave sources, we have studied the cross-correlation signal for the galaxy distribution shown by Model-II with $z_0= 0.5$. The colored bands shows the tomographic bins chosen in this analysis with $\Delta z=0.2$ for LIGO and $\Delta z=0.5$ for LISA.}
    \label{fig:dndz}
\end{figure*}
\subsection{Estimator}
The observed gravitational wave signal depends upon both gravitational wave strain from the source and the effect from the gravitational wave propagation, which can be expressed according to Eq. \eqref{gw-lensing}.  
The observed $D_L$ from the gravitational wave strain can be written as
\begin{align}\label{gw-len-est2}
\begin{split}
    D_L(\hat n)&= \frac{d_L (\hat n)}{1+ \kappa(\hat n)}+\epsilon_{gw}(\hat n),\\
    &= d_L (\hat n) (1- \kappa(\hat n)) + \epsilon_{gw}(\hat n) +O(\kappa^2),
\end{split}
\end{align}
where  $\epsilon_{gw}$ is the observational error from the gravitational wave observation with zero mean  {and $d_L= (c(1+z_s)/H_0)\int_0^{z_s} \frac{dz'}{\sqrt{\Omega_m(1+z')^3 + (1-\Omega_m)}}$}. In the weak lensing limit $\kappa_{gw}<<1$, which is used to obtain the second equation in Eq. \ref{gw-len-est2}. The luminosity distance can be inferred independently of the chirp mass using the relation derived by \citet{1986Natur.323..310S}
\begin{equation}\label{gw-dl-relation}
\begin{split}
D_L \propto & \frac{1}{\bar h(t) \tau \nu^2},\, 
\text{where } \tau \equiv  \bigg(\frac{d\nu/dt}{\nu}\bigg)^{-1} \propto \frac{\pi M_z^2}{(\pi M_z)^{11/3}\nu^{8/3}},\\& \text{and } \bar h(t) \propto\frac{M_z (\pi \nu_zM_z)^{2/3}}{d_L},
\end{split}
\end{equation}
where $\tau$ is the time-scale related to the change of the frequency of the gravitational wave signal and $\bar h$ is the gravitational wave strain averaged over detectors and source orientations. However, the source inclination angle is important to estimate the luminosity distance accurately \citep{2010ApJ...725..496N}. By using the two polarization states of the gravitational wave signal $h_+$ and $h_\times$, the degeneracy between the source inclination angle and  luminosity distance can be lifted \citet{2010ApJ...725..496N}. 

Eq. \eqref{gw-len-est2} shows that there exists a multiplicative bias to $\kappa_{gw}$ from $d_L$ . So, in order to remove this bias, we need to make an estimate of the true  luminosity distance for each sources. The estimate of true luminosity distance to the gravitational wave sources can be estimated using the redshift ($z_s$) of the source (from the EM counterparts \citep{Nissanke:2012dj}) and best-fit cosmological parameters (obtained using the data from other cosmological probes such as CMB, supernovae, etc.) in the relation $d^{es}_L= d_L(1+\epsilon_s)$, where $\epsilon_s$ is the error in the measurement of true $d_L$ due to redshift error and the error in best-fit cosmological parameters. 
Then the estimator for $\kappa_{gw}$ using the the ratio of $D_L$ and $d^{es}_L$ is
\begin{align}\label{gw-len-est4}
\begin{split}
    \mathcal{\hat D}_L (\hat n)&\equiv 1-\frac{D_L (\hat n)}{d^{\text{es}}_L(\hat n)} \\
&= 1- \frac{(1- \kappa_{gw} (\hat n))+ \frac{\epsilon_{gw}}{d_l}(\hat n)}{1+\epsilon_s(\hat n)}.
\end{split}
\end{align}
For $\epsilon_s <<1$, we can write the above equation as
\begin{align}\label{gw-len-est}
\begin{split}
\mathcal{\hat D}_L (\hat n) &= \kappa_{gw} (\hat n) - \frac{\epsilon_{gw}}{d_l}(\hat n) + \epsilon_s(\hat n) - \epsilon_s(\hat n) \kappa_{gw}(\hat n). 
\end{split}
\end{align}
This is feasible for the gravitational wave sources with electromagnetic counterparts, the first case we will discuss. For those gravitational wave sources that do not have electromagnetic counterparts,  eq. \ref{gw-len-est4} is the estimator of the lensing signal, which produces a multiplicative bias in estimation of the lensing potential $\kappa_{gw}$. We will return to this case below and discuss how to remove the bias.

\subsubsection{Gravitational wave events with electromagnetic counter part}
For the case when $\epsilon_s <<1$  the corresponding estimator for  gravitational wave lensing and galaxy lensing cross-correlations in  real space can be written as\footnote{The quantities with hat denote the estimator of the signal}
\begin{align}\label{gw-len-gal-len}
\begin{split}
\hat C^{\kappa_{gw}\kappa_{g}}(\hat n .\hat n')&= \langle \mathcal{\hat D}_L (\hat n) \kappa_{g} (\hat n)\rangle,\\
&=  \langle \kappa_{gw} (\hat n)  \kappa_{g} (\hat n')\rangle + \cancel{\langle \epsilon_s (\hat n)  \kappa_{g} (\hat n)\rangle} \\ &- \cancel{\langle \epsilon_s(\hat n)\rangle} \langle\kappa_{gw}(\hat n) \kappa_{g}(\hat n')\rangle,\\
\hat C^{\kappa_{gw}\delta}(\hat n .\hat n')&= \langle \mathcal{\hat D}_L (\hat n) \delta (\hat n) \rangle,\\
&=  \langle \kappa_{gw} (\hat n)  \delta (\hat n')\rangle + \cancel{\langle \epsilon_s (\hat n)  \delta (\hat n)\rangle} \\ &- \cancel{\langle \epsilon_s(\hat n)\rangle} \langle\kappa_{gw}(\hat n) \delta(\hat n')\rangle.
\end{split}
\end{align}
Here, the term $\epsilon_s$ is not correlated with the lensing and density field of the galaxy $\delta_g$ and $\kappa_{gw}$ and hence goes to zero on  averaging over a large number of sources. As a result, the estimator of the cross-correlation remains unbiased by any error in the luminosity distance measurement for the gravitational wave sources with electromagnetic counterpart. 

\subsubsection{Gravitational wave events without electromagnetic counterpart} 
Returning now to the case of gravitational wave sources without  electromagnetic counterparts, the error term $\epsilon_s \sim 1$. The corresponding cross-correlation signal can be written as
\begin{align}\label{gw-len-gal-len-noec}
\begin{split}
    \hat C^{\kappa_{gw}\kappa_{g}}(\hat n .\hat n')&= \langle\frac{\kappa_g(\hat n')\kappa_{gw}(\hat n)}{1+\epsilon_s(\hat n)}\rangle,\\
    &= C^{\kappa_g\kappa_{gw}}(\hat n.\hat n')\bigg\langle\frac{1}{1+\epsilon_s(\hat n)}\bigg\rangle,\\
    \hat C^{\kappa_{gw}\delta}(\hat n .\hat n')&= \langle\frac{\kappa_{gw}(\hat n)\delta(\hat n')}{1+\epsilon_s(\hat n)}\rangle,\\
    &= C^{\kappa_{gw}\delta}(\hat n.\hat n')\bigg\langle\frac{1}{1+\epsilon_s(\hat n)}\bigg\rangle,
\end{split}
\end{align}
The estimated cross-correlation signal can have a multiplicative bias if $\langle \epsilon_s \rangle \neq 0$ for sources without EM counterparts. However,
 for a large number of gravitational wave sources, all that is required to remove this bias is $N_{gw}(z)$, the redshift distribution of the gravitational wave sources. Using the clustering approach, redshifts of the gravitational wave sources can be obtained 
 using the methods described in \citep{Menard:2013aaa, PhysRevD.93.083511, Mukherjee:2018ebj}.
 
These estimators in the spherical harmonic basis \footnote{Any field $X(\hat n)$ can be expressed in the spherical harmonics basis as $X(\hat n)= \sum_{lm} X_{lm} Y_{lm}(\hat n)$} can be expressed as $\hat C^{\kappa_{gw}\kappa_{g}}_l=  \frac{1}{2l+1}\sum_m (\kappa_{gw})_{lm}  (\kappa_{g})_{lm}$ and $\hat C^{\kappa_{GW}\delta}_l=  \frac{1}{2l+1}\sum_m (\kappa_{gw})_{lm} \delta_{lm}$. The real space cross-correlation estimator given in Eq. \ref{gw-len-gal-len} is related to the spherical harmonic space cross-correlation estimator by the relation
 \begin{align}\label{gw-len-gal-len-sh}
\begin{split}
\hat{C}_\mathcal{XX'}(\hat n. \hat n')= \sum_l \bigg(\frac{2l+1}{4\pi}\bigg) P_l(\hat n. \hat n') \hat C^{XX'}_l,
\end{split}
\end{align}
where, $P_l(\hat n. \hat n') $ are the Legendre polynomials. 

\begin{figure*}
    \centering
    \includegraphics[trim={0 0 0 0cm}, clip, width=1\textwidth]{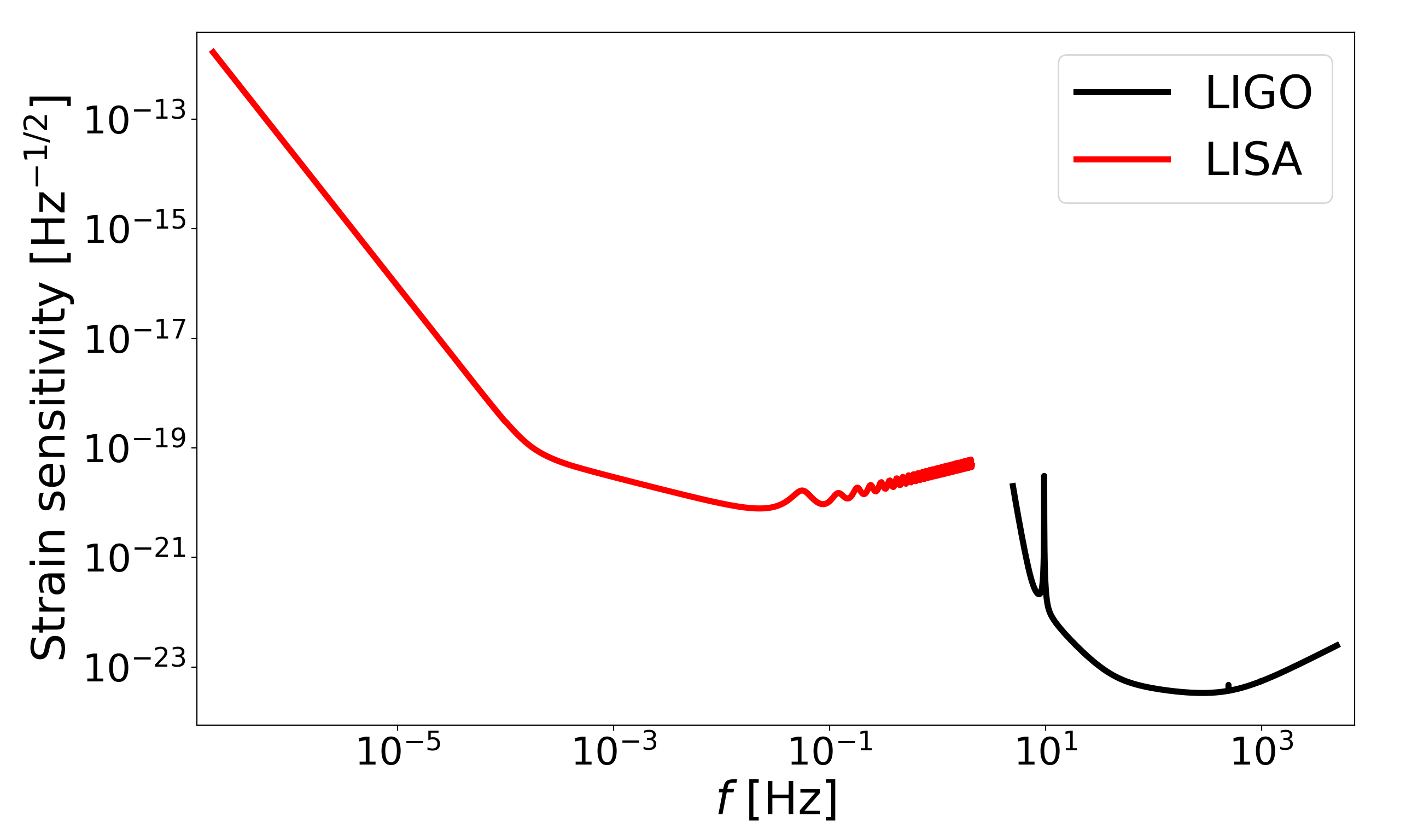}
    \caption{We plot the gravitational wave strain noise for Advanced-LIGO and LISA. The Advanced-LIGO noise is from the latest projected noise.  The LISA noise is plotted for one year of integration time with the LISA instrument specifications given in Table \ref{tab-1}.}
    \label{fig:gwnoise}
\end{figure*}

The variance of the cross-correlation signal can be written in terms of the available overlapping sky fraction of both the missions $f_{sky}$ as 
\begin{align}\label{gw-gal-noise-1}
\begin{split}
(\sigma^{gw-X}_l)^2  = \frac{1}{f_{\text{sky}}(2l+1)}& \bigg((C^{\kappa_\text{gw}\kappa_\text{gw}}_l + N_l^{\mathcal{DD}})(C^{{XX}}_l  + N^{XX}_l) \\& + (C^{\kappa_{gw}{X}}_l)^2\bigg),\\
\end{split}
\end{align}
where, $X \in g, \kappa_{g}$, $N^{XX}_l= \sigma_X^2/n_X$ corresponds to the shape noise ($\sigma_{\kappa_{g}}= 0.3$) and the galaxy shot noise ($\sigma_g= 1$) for $X=\kappa_{g}$ and $g$ respectively. $n_X$ denotes the number density of sources per sr.  
$N_l^{\mathcal{DD}}$ is the variance related to the estimation of the gravitational wave lensing signal, which can be written as
\begin{align}\label{gw-noise-1}
\begin{split}
N_l^{\mathcal{DD}} = \frac{4\pi}{N_{gw}}\bigg(\frac{\sigma^2_{d_l}}{d^2_l} + \frac{\sigma^2_{b}}{d^2_l}\bigg)e^{l^2\theta_{min}^2/8\ln 2},
\end{split}
\end{align}
where  $N_{gw}$ is the number of gravitational wave sources, $\theta_{min}$ is the sky localization area of the gravitational wave sources, $\sigma^2_{d_l}$ is the error in the luminosity distance due to the detector noise and $\sigma^2_{b}$ is the error due to the error in the gravitational wave source redshift and uncertain values of cosmological parameters.

The joint estimation of all the effects from the  propagation of the gravitational wave signal can be written in an unified frame work by combining the observable from the gravitational wave and the cosmic probes such as the galaxy surveys $\delta_g$ and $\kappa_g$ as 
\begin{equation}
\mathcal{O}=
\begin{pmatrix} 
\hat{D}_L \\
\hat \delta\\
\hat \kappa_{g} 
\end{pmatrix}     
\end{equation}
and the corresponding covariance matrix can be expressed as
\begin{align*}
\begin{split}
C\equiv\langle \mathcal{O}\mathcal{O}^\dagger \rangle =    
\begin{pmatrix} 
C_l^{D_LD_L}  & \mathbf{C_l^{\kappa_{gw}\delta}}  & \mathbf{C_l^{\kappa_{gw}\kappa_{g}}} \\
\mathbf{C_l^{\kappa_{gw}  \delta}} & C_l^{\delta\delta} & C_l^{\delta\kappa_{g}} \\
\mathbf{C_l^{\kappa_{gw}\kappa_{g}}}   & C_l^{\delta\kappa_{g}} & C_l^{\kappa_{g}\kappa_{g}}
\end{pmatrix}.
\end{split}
\end{align*}
Here, the terms in bold indicates the new cosmological probes which will be available from the cross-correlation of the gravitational wave signal with the galaxy field. The cross-correlation of the galaxy surveys with the stochastic gravitational wave signal is another probe to study the unresolved sources \citep{Mukherjee:2019oma}.

\section{Forecast for the measurement of lensing of gravitational wave}\label{forecast}
The measurement of the galaxy lensing/gravitational wave lensing and galaxy/gravitational wave lensing can be explored from upcoming missions. The large-scale structure of the Universe will be probed by future missions such as DESI \citet{Aghamousa:2016zmz}, EUCLID \citet{2010arXiv1001.0061R}, LSST \citet{2009arXiv0912.0201L}, SPHEREx \citet{Dore:2018kgp}, WFIRST \citep{Dore:2018smn},  having a wide sky coverage and access to galaxies up to high redshift ( about $z \sim 3$). For the large-scale structure surveys, we consider the survey setup with two different normalized redshift distributions according to Eq. \ref{dndzeq} with the value of $z_0= 0.2$ and $z_0=0.5$. We have considered two different sets of number density of galaxies as $n_{g}= 0.5$ per arcmin$^2$ for $z_0=0.2$ and $45$ per arcmin$^2$ for $z_0=0.5$ in this analysis. The redshift distribution of the large scale structure surveys overlaps with the the redshift distribution of the gravitational wave sources. 
\begin{figure*}
\centering
\subfloat[]{\includegraphics[trim={0.2cm 0.2cm 0.5cm 0.5cm},clip,width=0.6\textwidth]{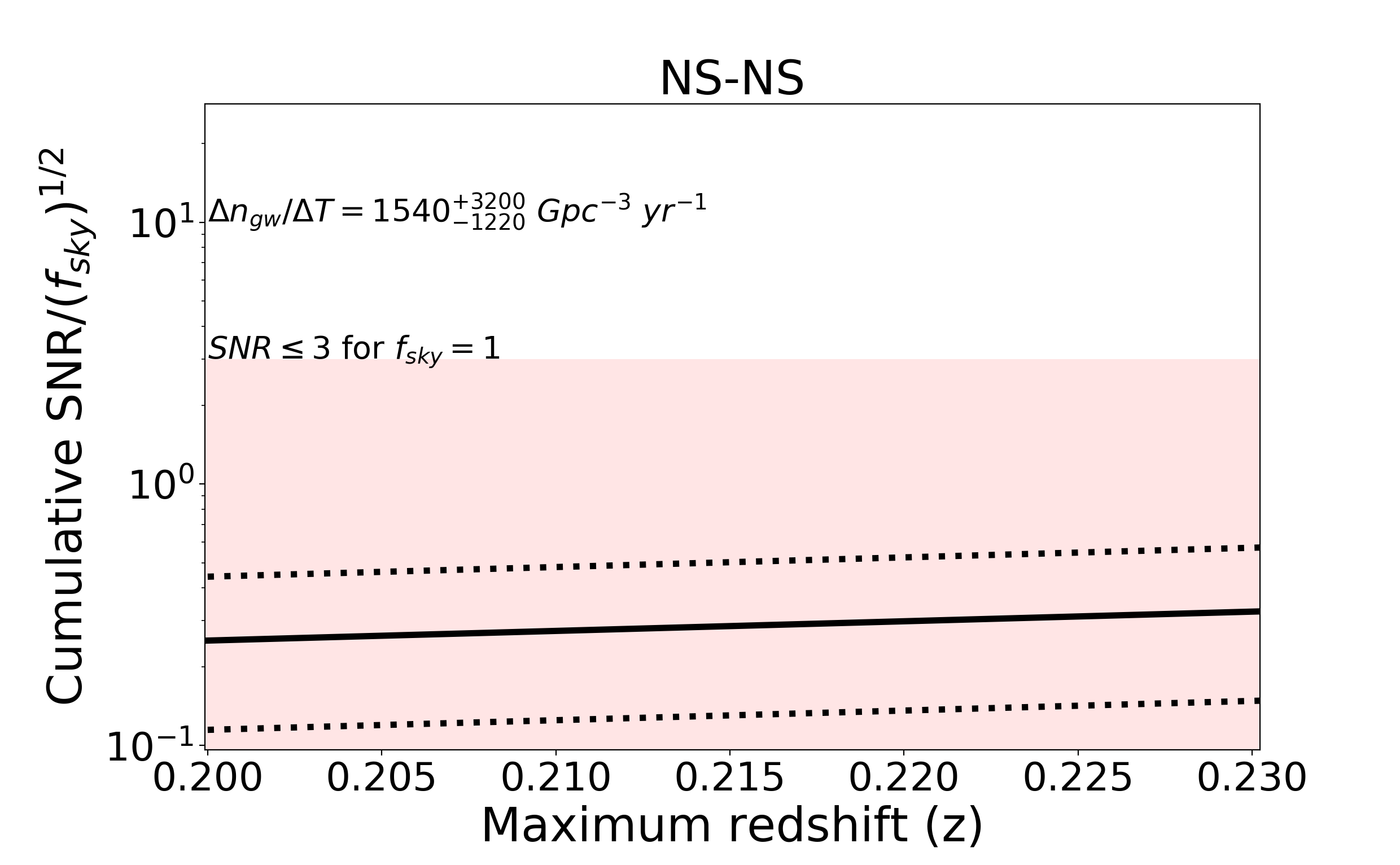}}\\
\subfloat[]{\includegraphics[trim={0.2cm 0.2cm 0.5cm 0.5cm},clip,width=0.6\textwidth]{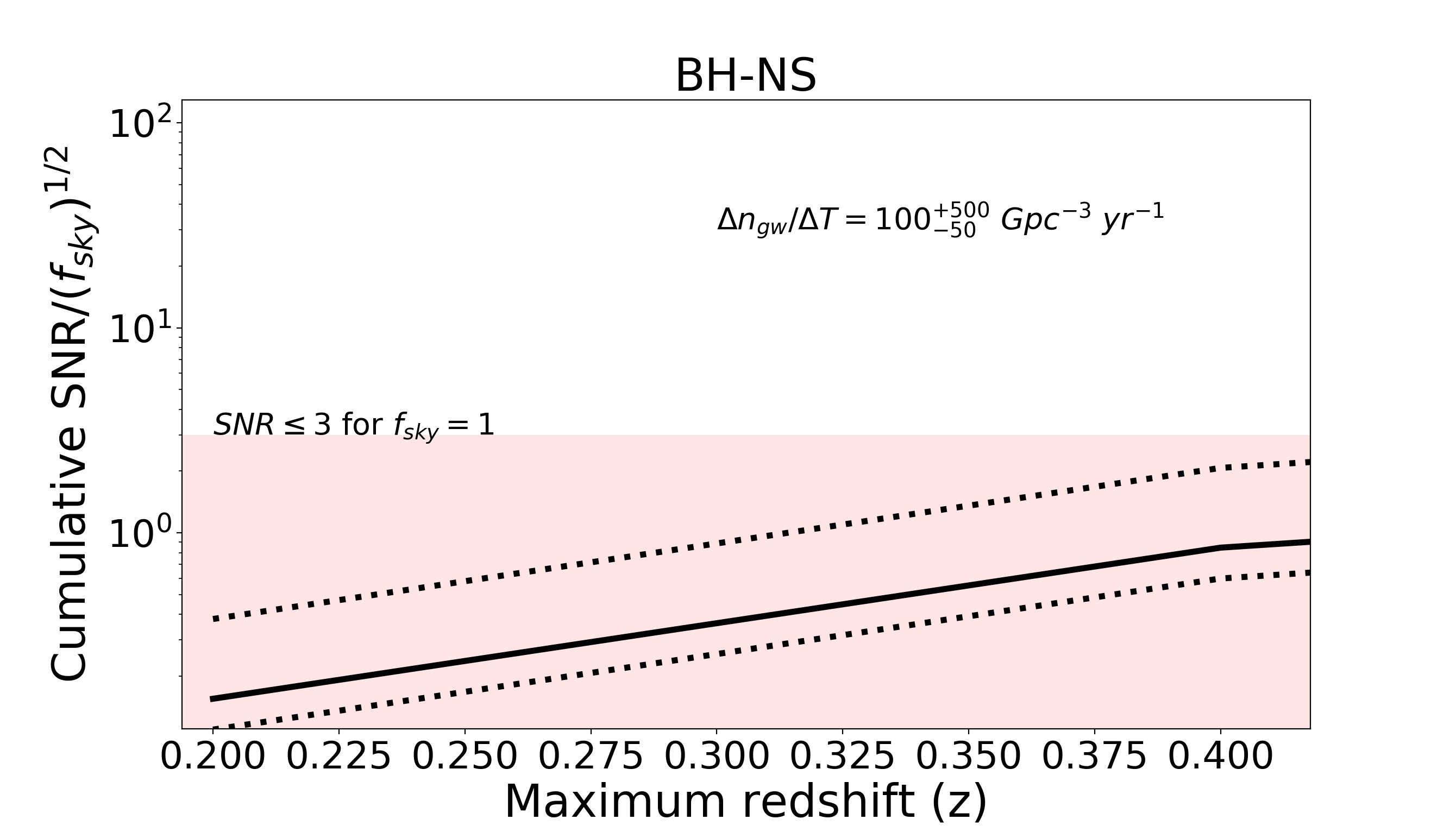}}\\
\subfloat[]{\includegraphics[trim={0.2cm 0.2cm 0.5cm 0.5cm},clip,width=0.6\textwidth]{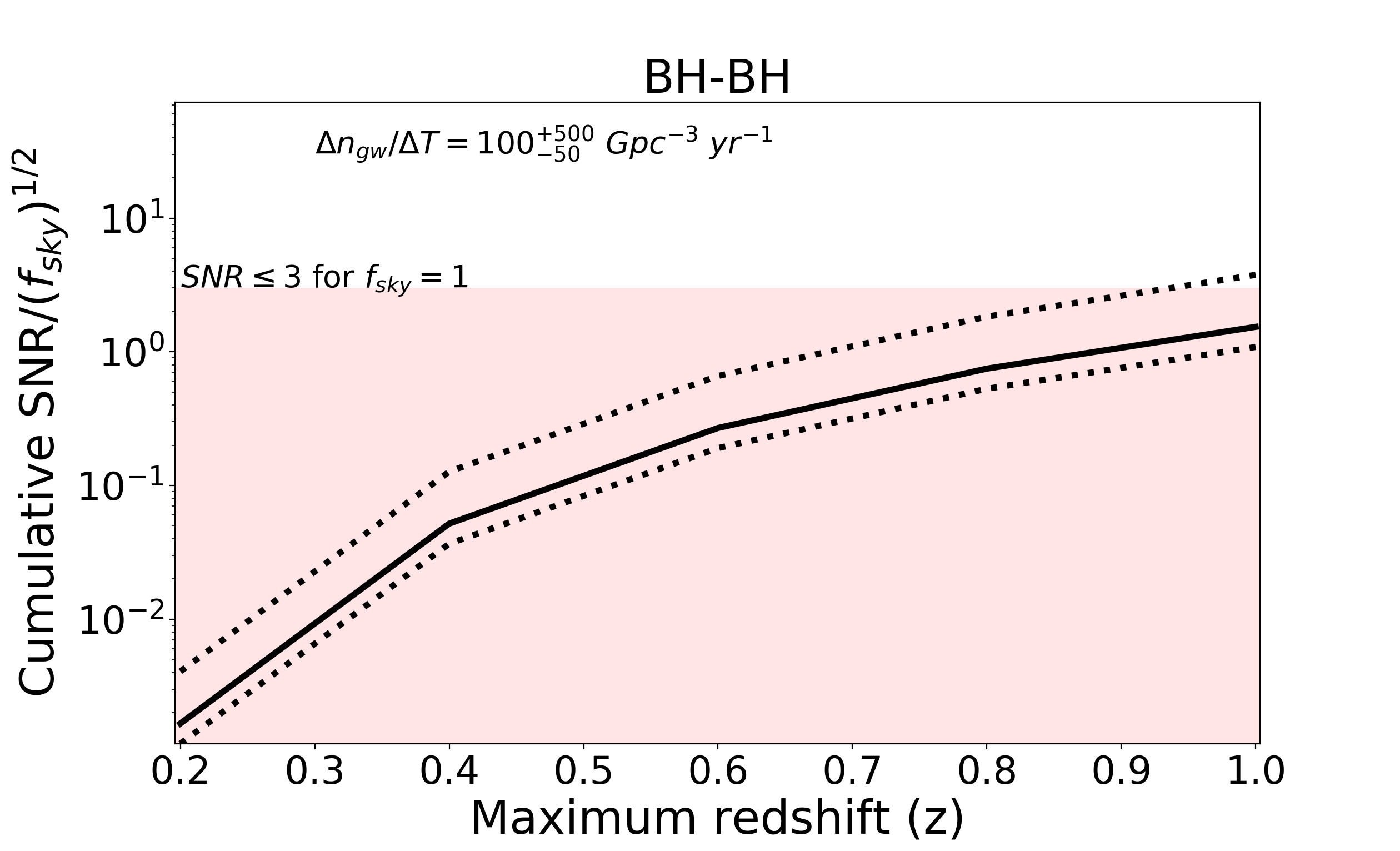}}
\caption{The cumulative SNR for the measurement of the cross-correlation between gravitational wave lensing and galaxy lensing $C_l^{\kappa_{gw}\kappa_{g}}$ for an observation time of $5$ years with the instrument noise of Advanced-LIGO. (a) For the NS-NS, the detection threshold of the gravitational wave sources above redshift of $0.2$ is going to be less than $10-\sigma$ and hence will not be identified as a LIGO event. (b) For BH-NS, we have considered here the case for a mass ratio of 20 with $M_{total}=31.5\, M_\odot$. (c) The BH-BH sources are considered for equal mass-ratios with $M_{total}= 60\, M_\odot$. The region shaded in pink indicates below $3-\sigma$ detection of the cosmological signal. The dotted-line indicates the upper and lower limits of the event rates, solid-line indicates the mean value of the event rates.}
\label{fig:bh_len}
\end{figure*}

\begin{figure*}
\centering
\subfloat[]{\includegraphics[trim={0cm 0.cm 0cm 0.cm},clip,width=0.6\textwidth]{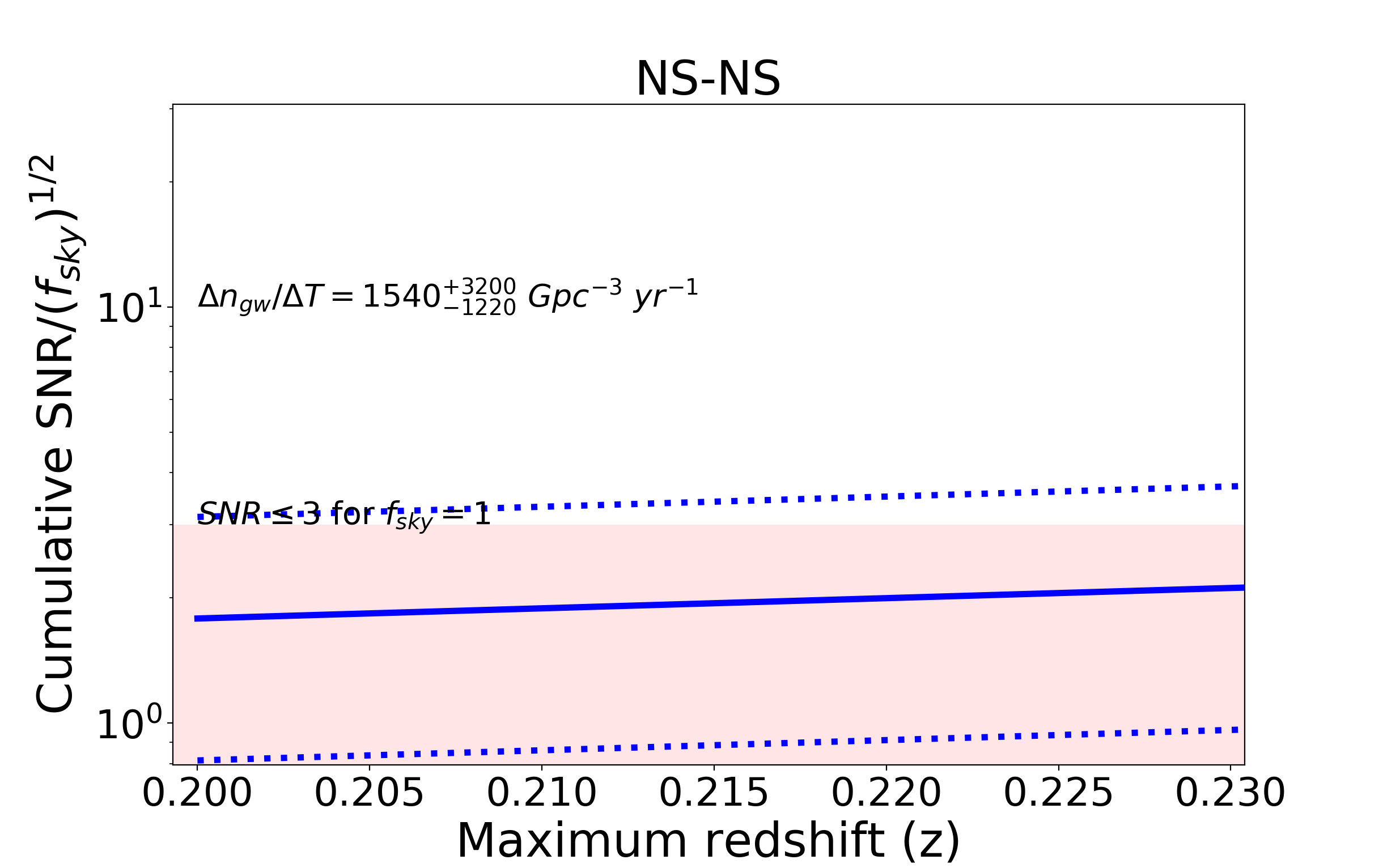}}\\
\subfloat[]{\includegraphics[trim={0cm 0.cm 0cm 0.cm},clip,width=0.6\textwidth]{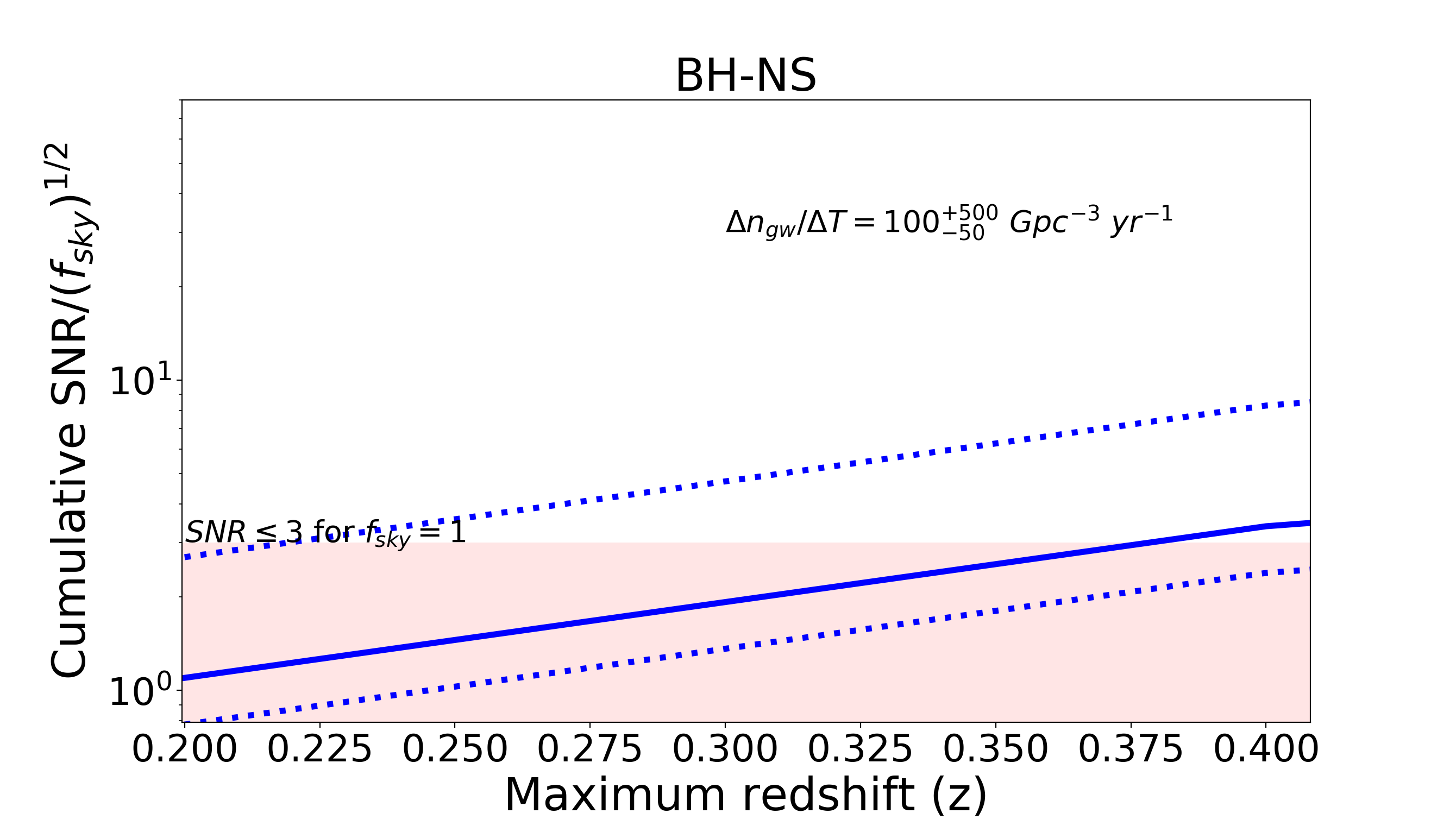}}\\
\subfloat[]{\includegraphics[trim={0cm 0.cm 0cm 0.cm},clip,width=0.6\textwidth]{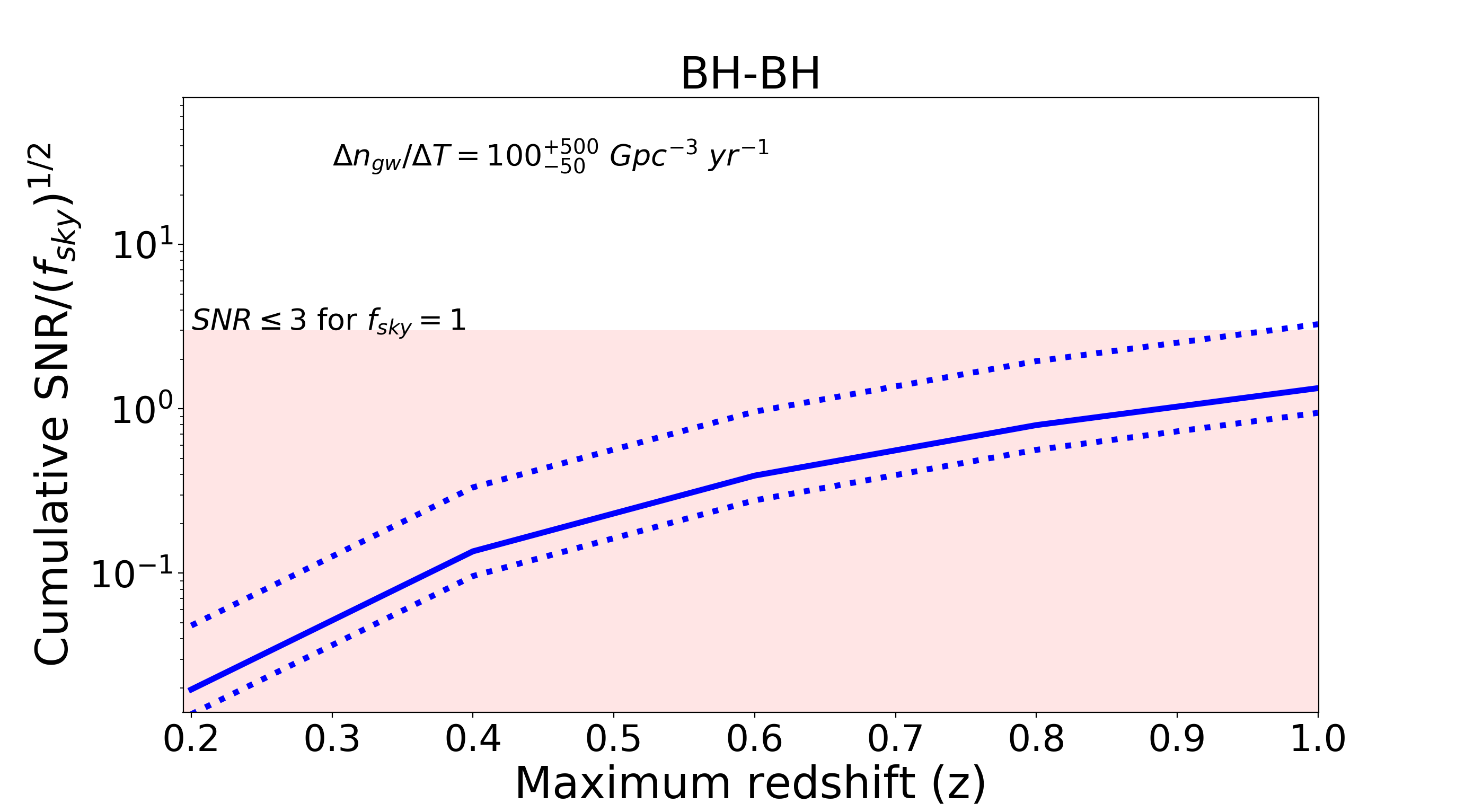}}
\caption{The cumulative SNR for the measurement of the cross-correlation between gravitational wave lensing and galaxy field $C_l^{\kappa_{gw}\delta}$ for an observation time of $5$ years with the instrument noise of Advanced-LIGO. (a) For the NS-NS, the detection threshold of the gravitational wave sources above a redshift of $0.2$ is going to be less than $10-\sigma$ and hence will not be identified as a LIGO event. (b) For BH-NS, we have considered here the case for a mass ratio of 20 with $M_{total}=31.5\, M_\odot$. (c) The BH-BH sources are considered for equal mass-ratio with $M_{total}= 60\, M_\odot$. The region shaded in pink indicates below $3-\sigma$ detection of the cosmological signal. The dotted-line indicates the upper and lower limits of the event rates, solid-line indicates the mean value of the event rates.}
\label{fig:bh_delta}
\end{figure*} 

For gravitational waves, we have considered the set-up for both  ground-based and  
space-based gravitational wave experiments such as Advanced-LIGO and LISA. For the gravitational wave detectors, we have considered the current instrument noise for Advanced-LIGO \footnote{\url{https://dcc.ligo.org/cgi-bin/DocDB/ShowDocument?.submit=Identifier&docid=T1800044&version=5}} and LISA as shown in Fig. \ref{fig:gwnoise}. The LISA noise curves are calculated using the online tool \footnote{\url{http://www.srl.caltech.edu/~shane/sensitivity/MakeCurve.html}} with the instrument specification provided in Table \ref{tab-1} \citep{PhysRevD.93.024003}.

The currently ongoing ground-based gravitational wave observatories such as Advanced-LIGO \citet{Evans:2016mbw} and VIRGO \citet{TheVirgo:2014hva}, and the future detectors such as KAGRA \citet{Akutsu:2018axf}, LIGO-India \citet{Unnikrishnan:2013qwa} will be joining these in the coming decade. These experiments will be capable of reaching up to a redshift of $z\sim 1$ and are going to detect the gravitational wave primarily from stellar origin binary black hole mergers  (BH-BH), black hole neutron star mergers (BH-NS) and binary neutron star mergers (NS-NS). The exact event rates of these binary mergers are not yet known and are predicted from the current observation of the gravitational wave events \citet{Abbott:2016nhf, PhysRevLett.119.161101}. 

We have taken the number of the sources of  gravitational waves as a free parameter for each of these binary species. Among these sources, BH-NSs and NS-NSs are expected to have an electromagnetic counterpart \citep{Bhattacharya:2018lmw,Nissanke:2012dj} and hence the redshift to these sources can be measured. We have considered the spectroscopic measurement of the source redshift for the electromagnetic counterpart from BH-NS and NS-NS. Sources such as BH-NSs and NS-NSs with electromagnetic counterparts can have an accurate estimation of the sky localization (less than one arcsec), which makes it possible to improve the cross-correlation signal with the LSS surveys. 
For the stellar mass BH-BHs, we do not expect any electromagnetic counterpart and as a result, the redshift of the source cannot be measured. For BH-BHs we have considered that the redshift of the source is completely unknown. For BH-BHs without any electromagnetic counterparts, we will be relying only on measurements from the gravitational wave signal and as a result, the sky localization error is going have relatively large errors compared to the case of BH-NS and NS-NS. For the combination of five detectors, we can expect about $10$ sq. deg sky localization error in the sources of the gravitational wave \citep{Nissanke:2011ax, Pankow:2018phc}. In this analysis, we have taken the angular resolution $\theta$ (or equivalently the maximum value of $l$, $l_{max}\propto 1/\Delta \theta$) as the free parameter and obtained the results for a few different choices. 
 \begin{table}
 \centering
\caption{Specifications of LISA instrument properties \citet{2017arXiv170200786A}  is used with $1$ year of integration time to produce Fig. \ref{fig:gwnoise} using the online tool \citet{PhysRevD.62.062001, PhysRevD.66.062001}.}
\label{tab-1} 
\vspace{0.5cm}
\begin{tabular}{||p{3.5cm}|p{4.5cm}|}
\hline 
\centering Specifications & \centering Values \tabularnewline
\hline
Armlength & \centering $2.5\times 10^9$ m\tabularnewline
\hline
Optics diameter & \centering $0.3$ m \tabularnewline
\hline
Wavelength & \centering $1064$ nm \tabularnewline
\hline
Laser power & \centering $2.0$ W \tabularnewline
\hline
Optical train efficiency & \centering $0.3$ \tabularnewline
\hline
Acceleration noise & \centering $3\times 10^{-15} \, \frac{m}{s^2\sqrt{Hz}}$ \tabularnewline
\hline
Position noise & \centering $2\times 10^{-9}\, \frac{m}{\sqrt{Hz}}$ \tabularnewline
\hline
Sensitivity noise floor  & \centering Position \\ \tabularnewline
\hline
Astrophysical Noise & \centering White dwarf 
\tabularnewline
\hline
\end{tabular}
\end{table}

Along with the ground-based detectors, space-based detectors such as LISA \citep{PhysRevD.93.024003, 2017arXiv170200786A, Smith:2019wny} will also be operational in the next decade and will be able to reach to  very high redshifts $z \sim 20$. These  experiments are going to explore  supermassive BH-BH systems over a wide range of masses $\sim 10^3- 10^7\, M_\odot$. Such sources are going to be observed in the frequency band $10^{-4}- 10^{-1}$ Hz for  time-scales of days to years before mergers of the binaries. The event rates for these sources are largely unknown and are mainly based on  studies made from simulations \citet{2007MNRAS.380.1533M}. For the heavier supermassive BH-BH systems such as $10^{7}\,M_\odot$, the event rates are going to be less than the case for the  intermediate supermassive BH-BH systems of mass $10^4\, M_\odot$, according to our current understanding of the formation of these binaries.  {We consider three different event rates for unit redshift bins, $50,\, 100$ and $300$ in this analysis for LISA. For the supermassive binary BHs, we can expect to see electromagnetic counterparts over a wide of electromagnetic frequency spectrum  \citep{2041-8205-752-1-L15, Haiman:2018brf,Palenzuela:2010nf,Farris:2014zjo,Gold:2014dta,Armitage:2002uu}. This is going to be useful for obtaining the redshift of  the gravitational wave sources by the electromagnetic follow up surveys.} 

In the future, the next-generation gravitational wave observatories will include Voyager \footnote{\url{https://dcc.ligo.org/public/0142/T1700231/003/T1700231-v3.pdf}}, which is going to achieve an instrument noise of $\sim 10^{-24}\,{\text{Hz}}^{-{1/2}}$. There are also proposals for ground-based observatories with larger baselines such as the Einstein Telescope \citep{Hild:2010id} and Cosmic Explorer \citep{Evans:2016mbw}. These observatories are capable of exploring stellar-mass BH-BHs, BH-NSs and NS-NSs up to a redshift about $z \sim 20$ for the frequency range $10-1000$ Hz. These observatories are going to have orders-of-magnitude improvement in the instrument noise over LIGO. As a result, our proposed methodology is going to be an exquisite probe of  the Universe for better understanding  gravity, dark matter and dark energy by combining the probes from gravitational wave and electromagnetic wave. In future work, we will estimate the prospects of this approach for the next-generation gravitational wave detectors.

The signal-to-noise ratio for the measurement of the cross-correlation of the gravitational wave lensing  by cross-correlating with the galaxy-lensing ($X=\kappa_{GW}$) and with the galaxy density field $X=\delta$ can be estimated using  $(S/N)^2= \sum_l^{l_{max}} \frac{C^{\kappa_{GW}X}_l}{(\sigma^{\kappa_{GW}X})^2}$.  Using the gravitational wave detector noise, we estimate the error in the luminosity distance $\sigma_{d_l}$ using a Fisher analysis by considering the two polarization states of the gravitational wave signal, 
  \begin{align}\label{fisher}
\begin{split}
F_{D_lD_l}= \sum_{+, \times}\frac{1}{2} \int_0^\infty 4 f^2 d\ln f \frac{\partial h_{+,\times}}{\partial D_l}\frac{1}{|h_n (f)|^2}\frac{\partial h_{+,\times}}{\partial D_l}.
\end{split}
\end{align}
The corresponding Cramer-Rao bound on the parameters is obtained as $\sigma_{D_l}= \sqrt{(\mathbf{F}^{-1}_{D_lD_l})}$. The noise due to lensing causes an error in the estimation of the luminosity distance. This can be written in terms of the fitting form as given by \citet{2010PhRvD..81l4046H}. The error in the redshift $\sigma_z=C(1+z)$ of the gravitational wave sources leads to an additional error in the estimation of the $d_l^{es}$ as $\sigma_{s}=\frac{\partial d_l}{\partial z}\sigma_z$.  We have taken $C=0.03$ for the cases with photo-metric measurements of the redshift, and $C=0$ for spectroscopic measurements of the redshift.

In Fig. \ref{fig:bh_len} and Fig. \ref{fig:bh_delta},  we show the cumulative SNR for the measurement of $C^{\kappa_{GW}\kappa_g}_l$ and  $C^{\kappa_{GW}\delta}_l$  signals for binary neutron star, black hole neutron star systems and binary black holes respectively. For binary neutron stars, current gravitational wave detectors will be able to detect individual objects with high SNR for redshifts $z<0.2$. As a result, we need to consider galaxy surveys which are going to probe the low redshift Universe such as DESI \citep{Aghamousa:2016zmz}. Using the galaxy distribution shown by the red colours in Fig. \ref{fig:dndz}, we estimate the cumulative SNR up to redshift $z=0.2$ for $l_{max}=3000$ and assuming spectroscopic measurements of the redshifts of the binary neutron stars. Detection is not possible for the currently considered event rate within five years of observation time from $C^{\kappa_{GW}\kappa_g}_l$. But for $C^{\kappa_{GW}\delta}_l$, we can expect a cumulative SNR of $3$ detection of the signal up to redshift $z=0.2$ for the highest event rates and the best sky-fraction scenario. For the redshift range $0.3>z>0.2$, LIGO will be detecting several very low SNR gravitational wave events. The cross-correlation signal with the low SNR gravitational wave events (which may not be considered as a detected event) and if the electromagnetic counterpart of these sources can be identified, then more than a $3$-$\sigma$ measurement is possible only from $C^{\kappa_{GW}\delta}_l$. However, this scenario is extremely rare and the detection of the signal beyond $z>0.2$ is unlikely. 

For the BH-NS systems, we have estimated the SNR for $M_{NS}= 1.5\, M_\odot$ and $M_{BH}= 30\, M_\odot$. These sources can be identified as individual events for redshifts below $0.5$ and we considered the low redshift galaxy survey as shown by the red coloured curve in Fig. \ref{fig:dndz} to estimate the cross-correlation signal between galaxy lensing and gravitational wave lensing. The measurability of the signal remains weak at nearly all the values of redshift for $C^{\kappa_{GW}\kappa_g}_l$ with $l_{max} =3000$. But the signal from  $C^{\kappa_{GW}\delta}_l$ can be detected with high SNR for redshift $z>0.35$ from the sources of the gravitational wave which will be detected by Advanced-LIGO-like detectors. BH-NS systems with the lower (or higher) mass of the black hole can be identified as  individual events only up to a lower (or higher) redshift, and the corresponding detection threshold will vary.     
The BH-NS systems are the best scenario for studying the cross-correlation signal as this can have an electromagnetic counterpart and can also be detected up to a high redshift than for the NS-NS systems. 

For the BH-BH systems, with $M_{total}= 60\, M_\odot$, gravitational wave events can be detected up to a redshift $z=1$. These sources are possibly not going to have electromagnetic counterparts and as a result, the redshifts of the objects will remain unknown.  {We have considered the redshift of the GW sources to be unknown to obtain the forecast for BBHs}. Along with the absence of the redshifts, these sources are going to have poor sky localization $\sim 10$ sq. degree even with the five detector network (LIGO-H, LIGO-L, Virgo, KAGRA, LIGO-India) \citep{Pankow:2018phc}. Using a minimum sky localization error of $10$ sq. degree and with  for the case with unknown redshift, we estimate the measurability of  $C^{\kappa_{GW}\kappa_g}_l$ and $C^{\kappa_{GW}\delta}_l$ for a galaxy survey shown by the black curve in Fig. \ref{fig:dndz} (which reaches up to redshift $z=1$). The measurability of the signal remains marginal for $5$ years of observation time with the current estimate of the LIGO events. The estimates of the mean signal can also be biased for a lesser number of gravitational wave samples, due to the unavailability of the true luminosity distance to the GW source (as shown in Eq. \ref{gw-len-est}). For a large number of the gravitational wave sources, clustering redshift of the gravitational wave source can be inferred by using the spatial correlations \citep{Menard:2013aaa, PhysRevD.93.083511,Mukherjee:2018ebj}.

Detection of more events is going to improve the projected SNR. In these estimates of the SNR for NS-NSs, BH-NSs, and BH-BHs, we have only considered the inspiral phase of the gravitational wave signal. Further improvement of the SNR is possible if we include merger and the ringdown phase of the gravitational wave strain. This will be considered in a future analysis including the complete waveform of the gravitational wave signal. 

\begin{figure*}
\centering
\subfloat[]{\includegraphics[trim={0cm 0.cm 0cm 0.cm},clip,width=0.7\textwidth]{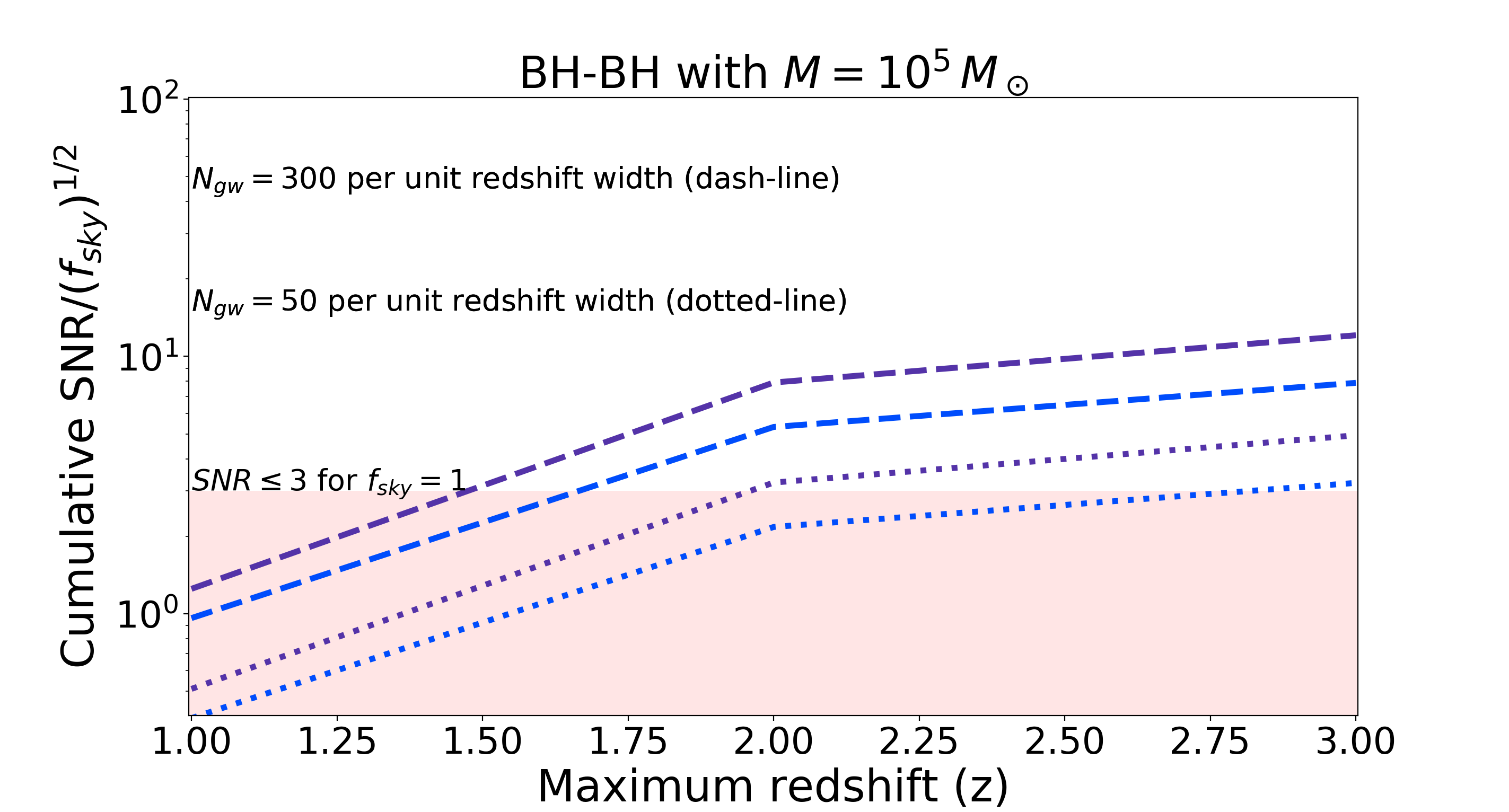}}\\
\subfloat[]{\includegraphics[trim={0cm 0.cm 0cm 0.cm},clip,width=0.7\textwidth]{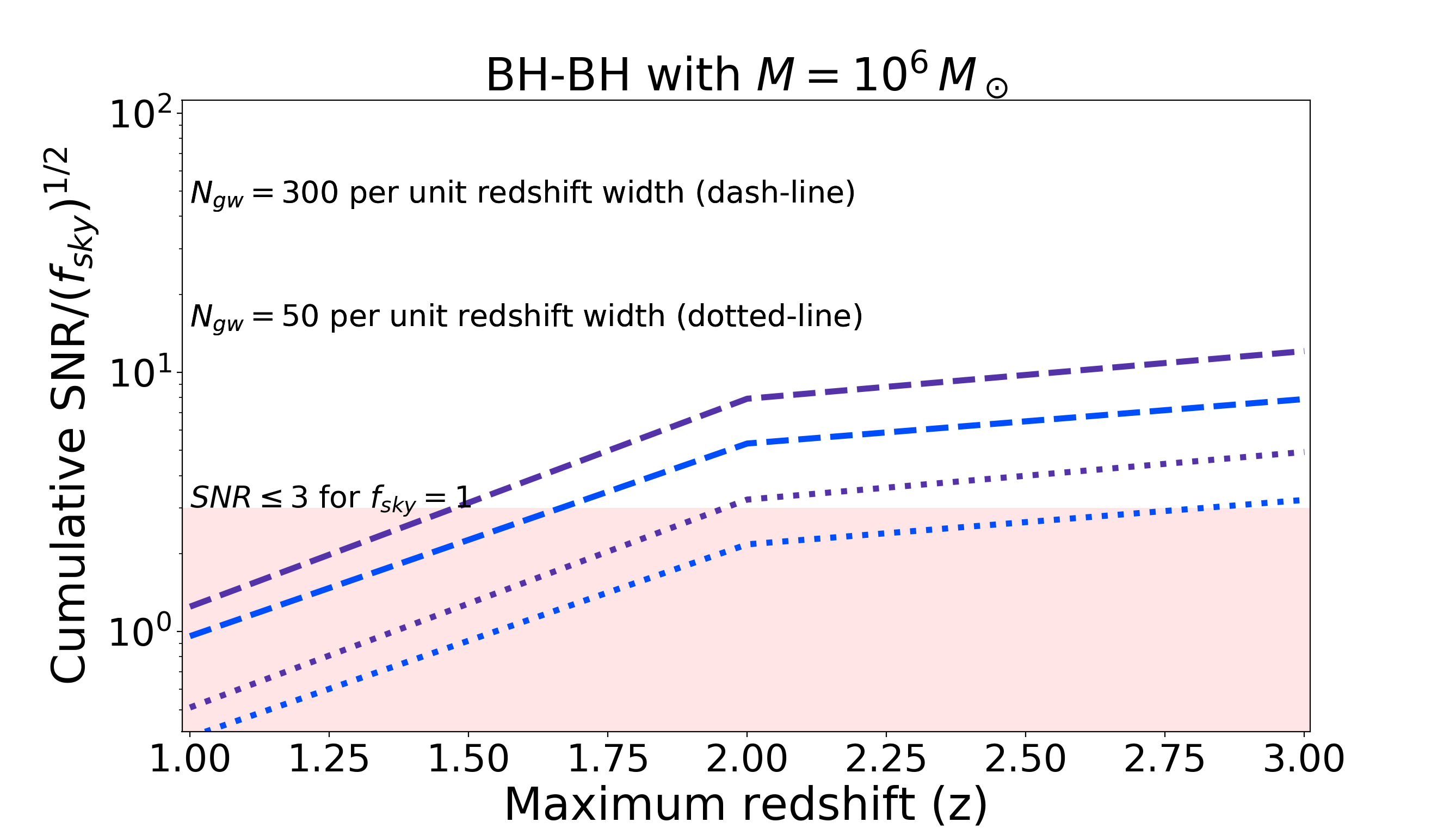}}\\
\subfloat[]{\includegraphics[trim={0cm 0.cm 0cm 0.cm},clip,width=0.7\textwidth]{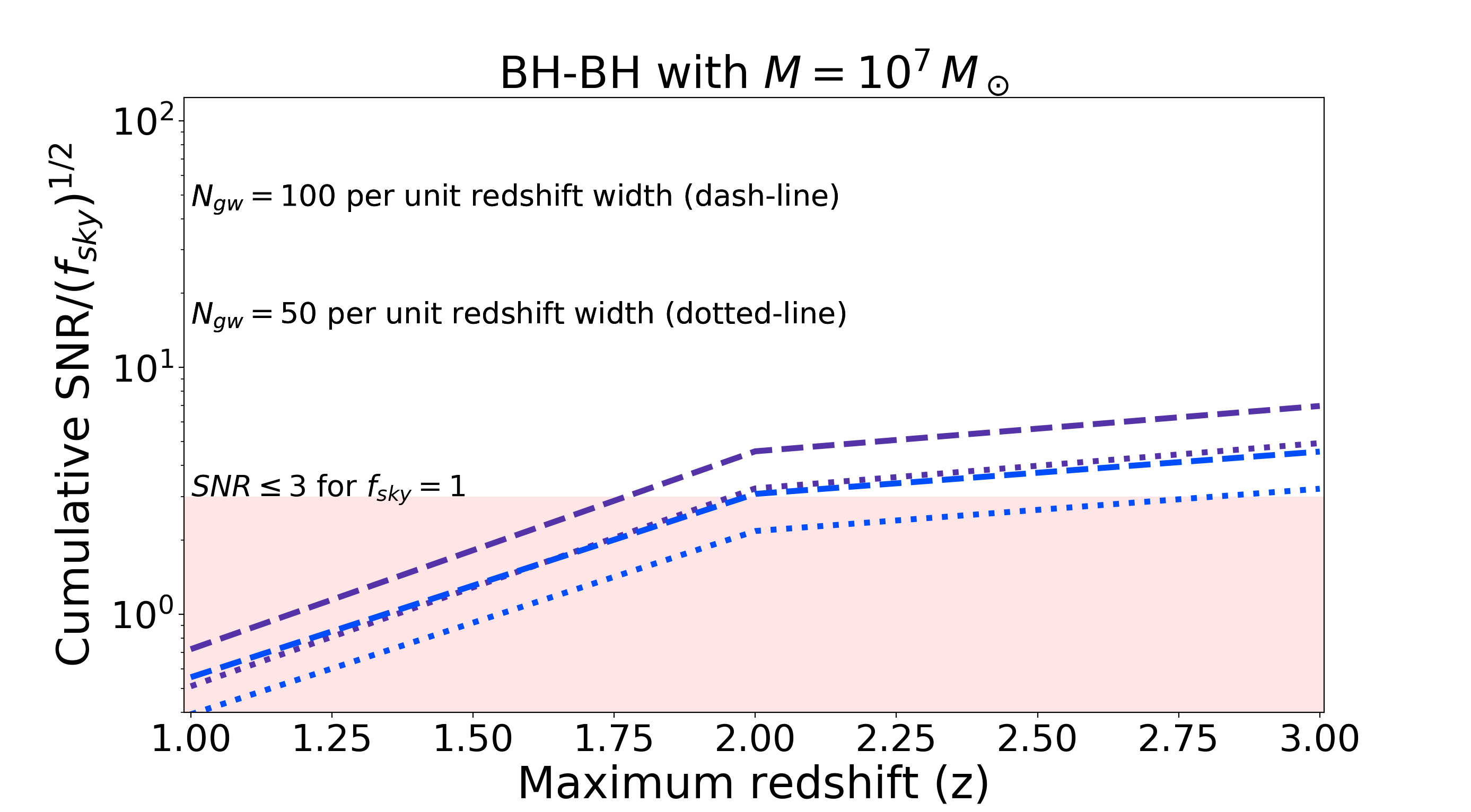}}
\caption{The cumulative SNR for the measurement of the cross-correlation between the gravitational wave lensing and galaxy lensing $C_l^{\kappa_{gw}\kappa_{g}}$ from LISA for an observation time of $4$ years for equal mass binary black holes with individual masses (a) $10^{5}$, (b) $10^6$ and (c) $10^{7}$. The region shaded in pink indicates below $3$-$\sigma$ detection of the cosmological signal. The projected SNRs are obtained for two sky localization  errors (i) $\Delta \theta= 1$ deg (shown in blue) and (ii) $\Delta \theta= 0.2$ deg (shown in purple).}
\label{fig:bh_len_lisa}
\end{figure*}

\begin{figure*}
\centering
\subfloat[]{\includegraphics[trim={0cm 0.cm 0cm 0cm},clip,width=0.65\textwidth]{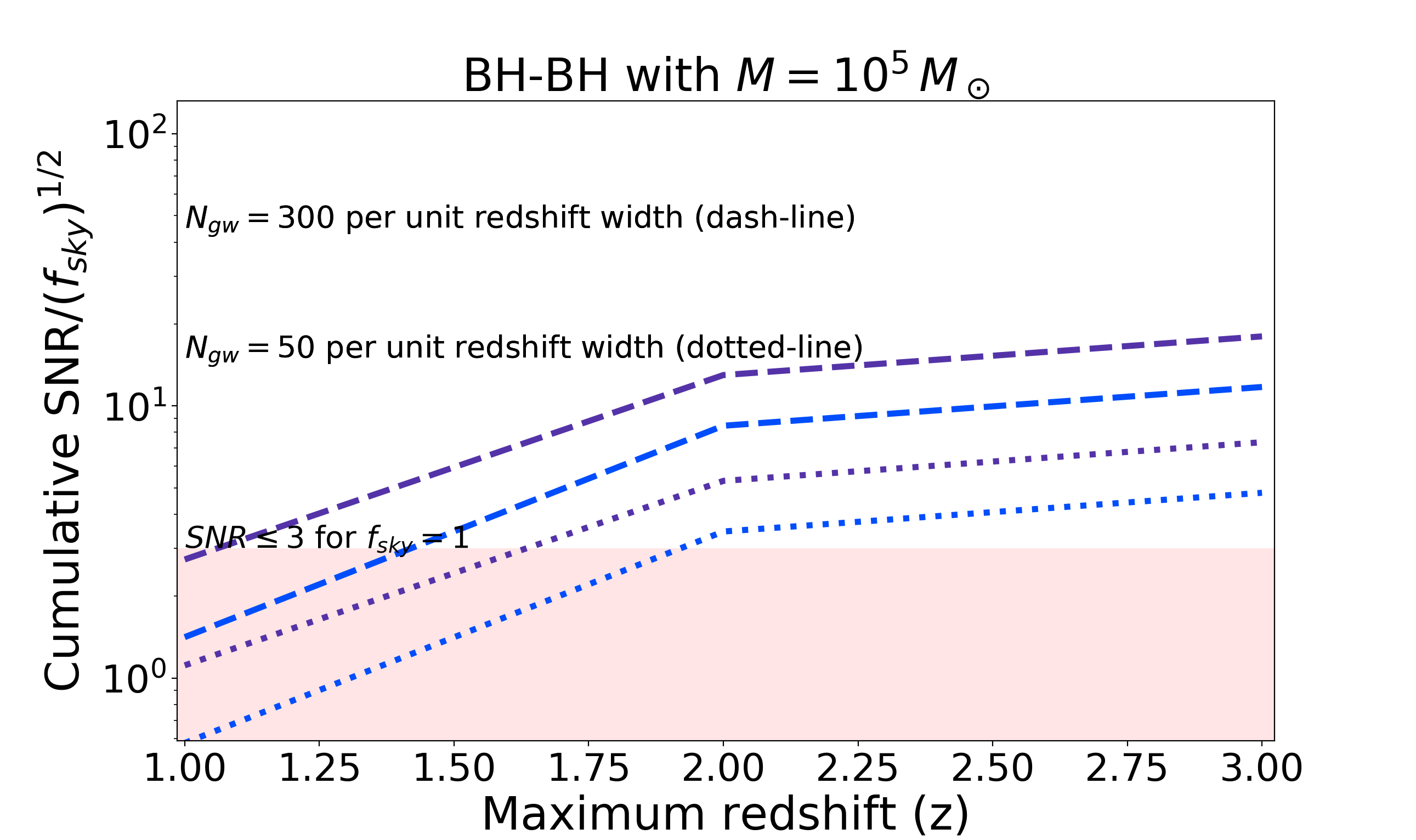}}\\
\subfloat[]{\includegraphics[trim={0cm 0.cm 0cm 0.cm},clip,width=0.65\textwidth]{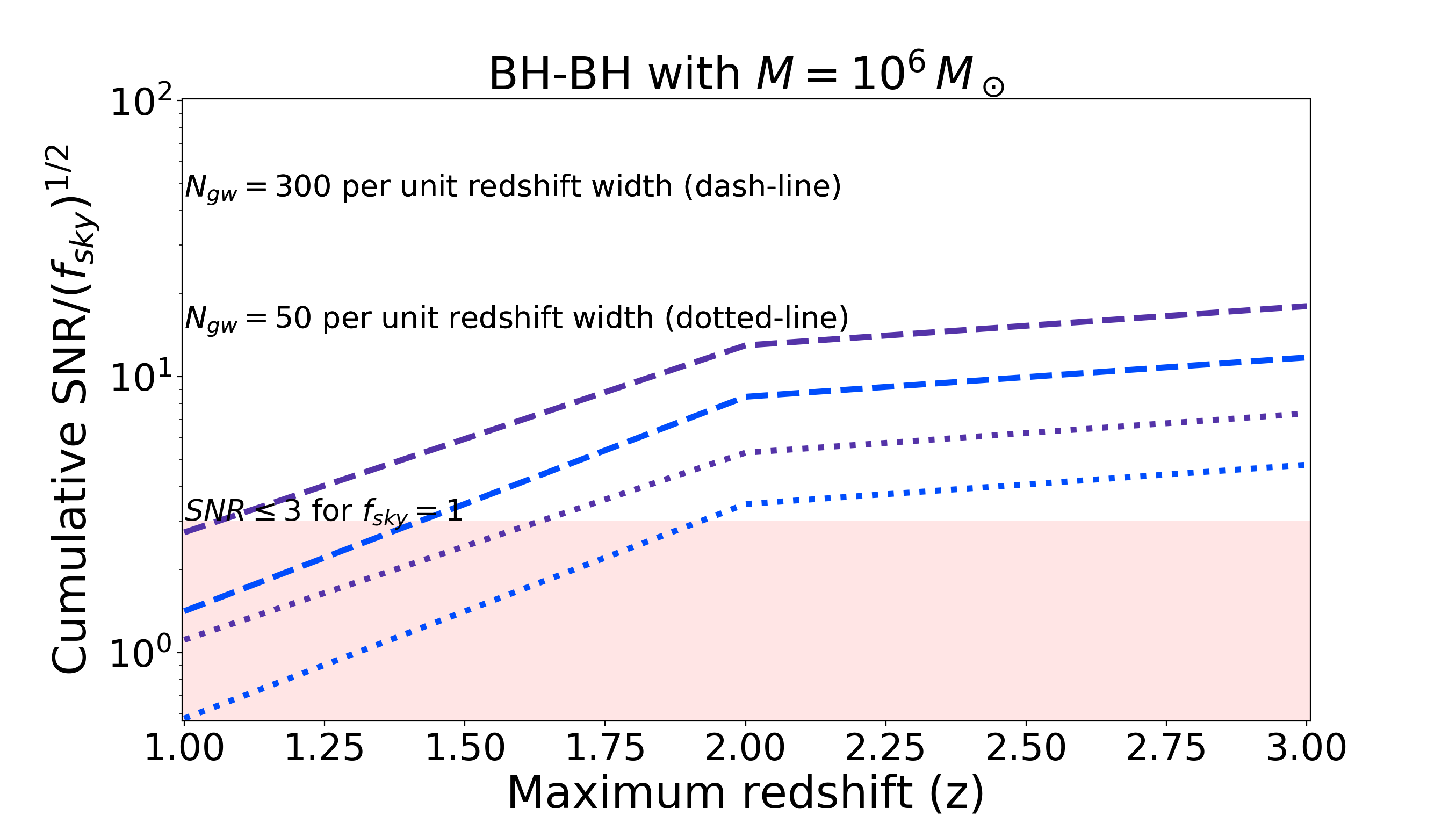}}\\
\subfloat[]{\includegraphics[trim={0cm 0.cm 0cm 0.cm},clip,width=0.65\textwidth]{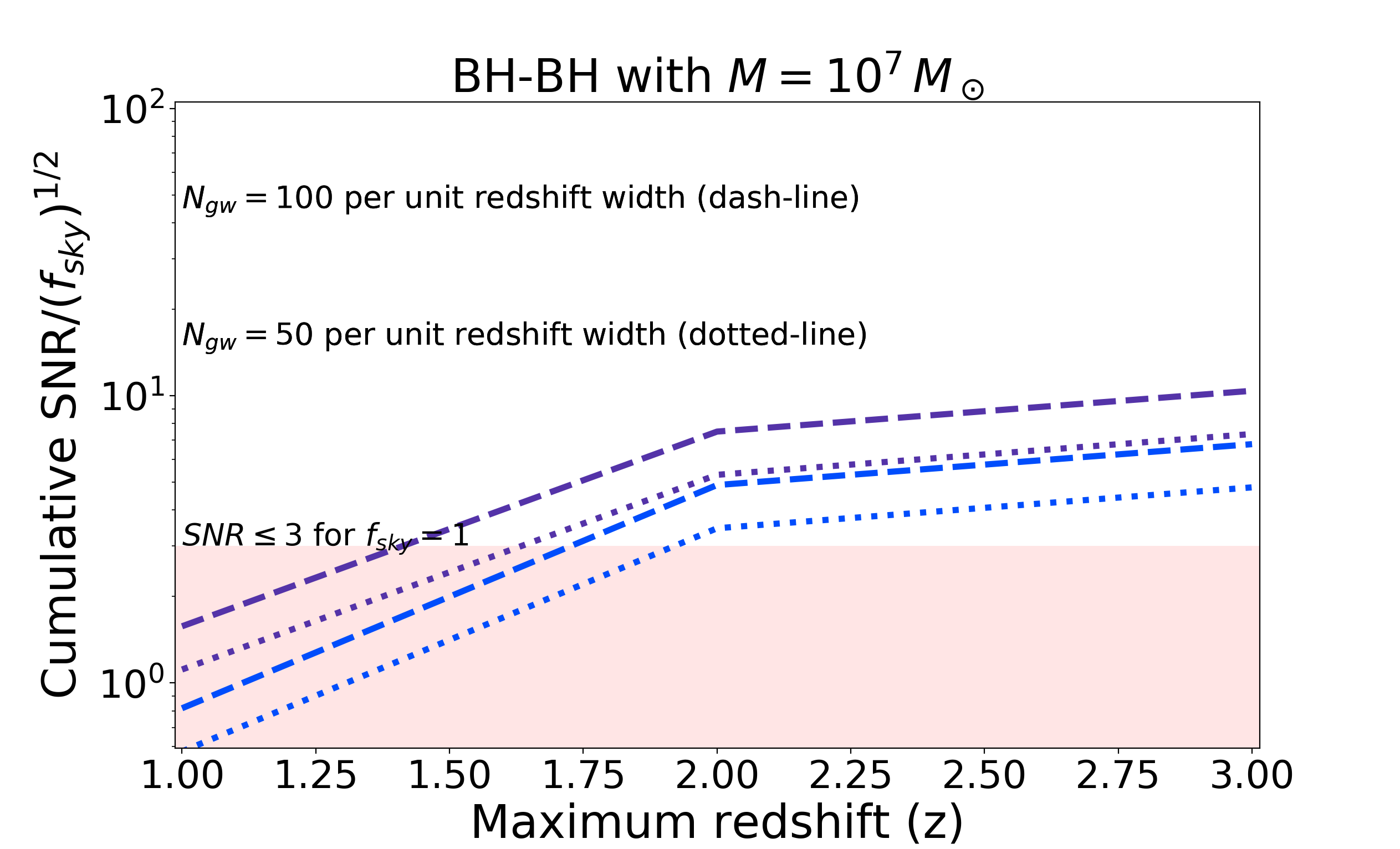}}\\
\caption{The cumulative SNR for the measurement of the cross-correlation between the gravitational wave lensing and galaxy field $C_l^{\kappa_{gw}\delta}$ from LISA for an observation time of $4$ years for equal mass binary black holes with individual masses (a) $10^{5}$, (b) $10^6$ and (c) $10^{7}$. The region shaded in pink indicates below $3$-$\sigma$ detection of the cosmological signal. The projected SNRs are obtained for two sky localization  errors (i) $\Delta \theta= 1$ deg (shown in blue) and (ii) $\Delta \theta= 0.2$ deg (shown in purple).}
\label{fig:bh_delta_lisa}
\end{figure*}

For the space-based gravitational wave detectors such as LISA, the gravitational wave events are going to be detected up to a redshift about $z\sim 20$ \citet{2017arXiv170200786A}. This enables us to probe the cosmological signal up to a high redshift by cross-correlating with  galaxy surveys which reach up to a high redshift. Using the redshift distribution of galaxies shown by the blue curve in Fig. \ref{fig:dndz} which mimics the surveys such as LSST, we estimate the SNR for the measurability of $C^{\kappa_{GW}\kappa_g}_l$ and $C^{\kappa_{GW}\delta}_l$ for different masses of the BH-BHs ranging from $10^4-10^7$. In this analysis, we have considered a maximum time period of one year of inspiral phase before the merger. The merger and the ringdown phase for these sources are not considered in the analysis. 
For the LISA BH-BH sources, we have considered a photometric error of $\sigma_z/ (1+z) = 0.03$ in the determination of the redshift of gravitational wave sources from the electromagnetic counterpart. Measurement of the electromagnetic counterpart is also going to be useful to reduce the sky localization error of the source of the gravitational wave and hence improves the detection of the cross-correlation signal. For two different values of $l_{max}$, $l_{max}=200$ and $l_{max}=1000$, we estimate the SNR for the measurement of $C^{\kappa_{GW}\kappa_g}_l$ and $C^{\kappa_{GW}\delta}_l$ signal for four years of observation time and show the corresponding estimates in Fig. \ref{fig:bh_len_lisa} and Fig. \ref{fig:bh_delta_lisa} respectively. As can be seen from these figures, the discovery space from synergy between LISA and LSST is promising for each mass window of BH-BHs. These SNRs are going to improve with the inclusion of the merger and the ring down phase of the gravitational wave signal. Further improvement in the SNR is possible with the increase in the LISA observation time to ten years, increase in the number of gravitational wave event rates and also with galaxy surveys which can go up to higher redshift. By combining all the high redshift probes such as  EUCLID \citet{2010arXiv1001.0061R}, LSST \citet{2009arXiv0912.0201L}, SPHEREx \citet{Dore:2018kgp} and WFIRST \citet{Dore:2018smn}, the detectability of the signal is going to improve significantly. At redshifts higher than $z>3$, the temperature and polarization anisotropies of the CMB is going to be a powerful probe to measure the lensing signal to even higher redshift as shown by \citep{Mukherjee:2019wfw}. Using the cross-correlation between CMB-lensing and gravitational wave lensing, we are going to have higher SNR measurements from redshift $z>3$ than possible from the galaxy surveys with the redshift distributions considered in this analysis (Fig. \ref{fig:dndz}).
Complementary information from cross-correlation with intensity mapping of the 21-cm line is another avenue to be explored in future work.

\section{Conclusions}\label{conclusion}
The general theory of relativity provides a unique solution to the propagation of gravitational waves which depends upon the underlying spacetime metric. Gravitational waves according to the theory of general relativity are massless and should propagate along null geodesics with the speed of light. As a result, in the framework of the standard model of cosmology, gravitational wave should propagate through the perturbed FLRW metric where the perturbations arise due to the spatial fluctuations in the matter density in every position. As a result of this metric perturbation, the gravitational wave strain is distorted.    

Study of these distortions in the gravitational wave strain while propagating through the perturbed metric can be a direct observational probe of several  aspects of fundamental physics, which include (i) the validation of general relativity from the propagation of gravitational wave signals and from the connection between  metric perturbations and matter perturbations, (ii) properties of dark energy and its effect on gravitational waves, (iii) gravitational effects of dark matter on gravitational waves and  evolution with cosmic redshift, (iv) testing the equivalence  principle from the multi-frequency character of the gravitational wave signal, (v) equivalence between the electromagnetic sector and the gravitational sector over the cosmic time-scale. 

The coming decade with  ongoing and upcoming gravitational wave observatories makes it possible to study these aspects and opens a new window in observational cosmology.  The astrophysical gravitational wave signals which are emitted from compact objects of different types, such as white dwarfs, neutron stars and black holes, span a vast  range of masses, gravitational wave frequencies and cosmological redshifts.   All of these sources are transients and are going to be frequent. As a result, they can probe the underlying theory of gravity and the statistical properties of the cosmological perturbations present in the Universe multiple times from different sources. As a result, all the above-mentioned aspects of fundamental physics can be studied using transient astrophysical gravitational wave sources.

In this paper, we have explored one of the important effects of matter perturbations, the weak lensing of the gravitational wave strain, and propose an avenue to explore this signal using cross-correlations with other cosmological probes such as  galaxy surveys.  The measurement of the lensing signal leads to a measurement of the evolution of the gravitational potential over a wide range of cosmic redshifts and also probes the connection of the matter perturbation with the gravitational potential. The joint study of cosmic density field and gravitational wave propagation in spacetime promises to be a powerful avenue towards a better understanding  of the theory of gravity and  the dark sector (dark matter and dark energy) of the Universe.

Several alternative theories of gravity and models of dark energy can be distinguished from the evolution of the cosmic structures \citep{Carroll:2006jn, Bean:2006up,Hu:2007pj,Schmidt:2008hc, Silvestri:2013ne, Baker:2014zba, Baker:2015bva, Slosar:2019flp} and also from the propagation of gravitational wave in  spacetime \citep{Cardoso:2002pa, Saltas:2014dha, Nishizawa:2017nef}. As a result, the studies of the cross-correlations between the gravitational wave signal and cosmic structures should result in a more comprehensive understanding of the theory of gravity and dark energy. The imprint of the lensing potential is also a direct probe of the dark matter distribution in the Universe, and  measurement of this via gravitational wave is going to explore the gravitational effects of dark matter on gravitational waves. The measurement of the gravitational lensing signal from  gravitational waves with a  frequency ranging from $10^{-4}$ to $10^{3}$ Hz will probe the universality of gravity on a wide range of energy scales and lead to a direct probe of the equivalence principle. Finally, this new avenue is also going to illuminate several hitherto unexplored windows which can be studied from the synergies between  electromagnetic waves and gravitational waves. 

Ground based gravitational wave detectors such as LIGO-US \citet{Evans:2016mbw} , Virgo \citet{TheVirgo:2014hva}, KAGRA \citet{Akutsu:2018axf}, LIGO-India \citet{Unnikrishnan:2013qwa} and space based gravitational wave detector such as LISA \citet{PhysRevD.93.024003, 2017arXiv170200786A} are going to probe a large range of cosmic  redshift $0<z\lesssim 20$ with gravitational wave sources such as NS-NS, BH-NS, stellar mass BH-BH and supermassive BH-BH. By cross-correlating with the data from  upcoming galaxy surveys such as DESI \citep{Aghamousa:2016zmz}, EUCLID \citep{2010arXiv1001.0061R}, LSST \citep{2009arXiv0912.0201L}, SPHEREx \citep{Dore:2018kgp}, WFIRST \citep{Dore:2018smn},  we will be able to obtain a high SNR detection of the weak lensing signal with these surveys, as shown in Fig. \ref{fig:bh_len}-\ref{fig:bh_delta_lisa}. The measurability is best for the sources with electromagnetic counterparts such NS-NS, BH-NS and super massive BH-BHs. For stellar mass black hole binaries, we do not expect electromagnetic counterparts, and hence poorly determined source redshifts. As a result, the measurability of the weak lensing signal is going to be marginal from stellar mass BH-BHs. 

We find that the most promising avenue with  Advanced-LIGO will be the cross-correlation of lensed black hole neutron star mergers and the galaxy overdensity $C^{\kappa_{GW}\delta}_l$. The contributions from binary neutron stars and binary black holes are marginal and only improve if the event rates are larger than considered in this analysis. However, with future surveys such as Cosmic Explorer and Einstein Telescope, we can explore the spatial cross-correlation of binary black holes and galaxies in order to obtain the clustering redshift \citep{Menard:2013aaa} of the sources \citep{PhysRevD.93.083511}. This can result in an improvement in the correlation signal. By fitting the shape of the correlation function between the spatial position of binary black holes and underlying galaxies, an accurate estimate of the source redshift is possible \citep{Mukherjee:2018ebj} for a large sample of binary black holes. The gravitational wave sources from LISA are going to provide a high SNR measurement from both $C^{\kappa_{GW}\kappa_g}_l$ and $C^{\kappa_{GW}\delta}_l$  over a vast range of gravitational wave source masses $10^4\, M_\odot-10^7\, M_\odot$ by cross-correlating with the large scale structure surveys with the access to high cosmological redshift. In future work, we will present a joint Bayesian framework for the combined analysis of the gravitational wave data and cosmological data. For this, we  use the full gravitational wave waveforms including the inspiral, merger and the ring-down phases, instead of only the inspiral phase  considered in this analysis. Inclusion of the merger and the ringdown phases is going to improve the SNR presented in this paper.    

In summary, we point out a new window for the emerging paradigm of multi-messenger observational cosmology with transient sources. This window will exploit both gravitational wave and electromagnetic waves in order to test the propagation and lensing of gravitational waves, potentially leading to high precision tests of gravity on cosmological scales. In addition to testing gravity in new ways, we expect this probe to be useful for developing an improved understanding of the standard model of cosmology. All the electromagnetic observational probes of the Universe point to additional matter content beyond  baryonic matter,  entirely inferred through its gravitational effects. The cause of these effects is associated in the standard model of cosmology with a cold dark matter component, which is neither understood theoretically, nor  directly  detected in  particle physics experiments. We show in this paper that gravitational waves will bring a new and independent window to confirm the existence of dark matter in the Universe and exploit the gravitational effect of dark matter on gravitational waves as a new and complementary probe on its physical nature. The detailed potential for constraints on the theoretical models from this new observational window, enabled by advances in gravitational wave detector capabilities and a new generation of galaxy redshift surveys, remain to be determined.

\section*{Acknowledgement}
S. M. would like to thank Karim Benabed, Luc Blanchet, Sukanta Bose, Neal Dalal, Guillaume Faye,  Zoltan Haiman, David Spergel, and Samaya Nissanke for useful discussions. The results of this analysis are obtained using the Horizon cluster hosted by Institut d'Astrophysique de Paris. We thank Stephane Rouberol for smoothly running the Horizon cluster. The Flatiron Institute is supported by the Simons Foundation. The work of SM and BDW is also supported by the Labex ILP (reference ANR-10-LABX-63) part of the Idex SUPER, and received financial state aid managed by the Agence Nationale de la Recherche, as part of the programme Investissements d'avenir under the reference ANR-11-IDEX-0004-02. The work of BDW is also supported by the grant ANR-16-CE23-0002. BDW thanks the CCPP at NYU for hospitality during the completion of this work. In this analysis, we have used the  following packages: IPython \citep{PER-GRA:2007}, Matplotlib \citep{Hunter:2007},  NumPy \citep{2011CSE....13b..22V}, and SciPy \citep{scipy}.

\bibliographystyle{mnras}
\bibliography{LSS-GW}
\label{lastpage}

\end{document}